\documentclass[12pt,preprint]{aastex}
\slugcomment{Submitted to ApJ.}

\newcommand{\units}{\quad\mathrm{erg\,cm^{-3}\,s^{-1}}}
\newcommand{\dvd}{\delta v^{-1}}

\def\crossarrow{%
\mathop{\vcenter{\hbox{\ooalign{\raise -0.1ex%
        \hbox{$\hbox{$\longrightarrow$\hspace*{.05em}}$}\cr\cr%
                $\hbox{\hspace*{.6em}/}$}}}}}%

\begin{document}

\title{Carbon isotope fractionation in protoplanetary disks}

\author{Paul M. Woods\altaffilmark{1} and Karen Willacy} \affil{Jet
  Propulsion Laboratory, California Institute of Technology,\\ 4800
  Oak Grove Drive, Pasadena, California, 91109, USA.}
\email{paul.woods@manchester.ac.uk} \altaffiltext{1}{Present address:
  Jodrell Bank Centre for Astrophysics, Alan Turing Building, The
  University of Manchester, Oxford Road, Manchester M13 9PL, UK}

\begin{abstract}
We investigate the gas-phase and grain-surface chemistry in the inner
30\,AU of a typical protoplanetary disk using a new model which
calculates the gas temperature by solving the gas heating and cooling
balance and which has an improved treatment of the UV radiation
field. We discuss inner-disk chemistry in general, obtaining excellent
agreement with recent observations which have probed the material in
the inner regions of protoplanetary disks. We also apply our model to
study the isotopic fractionation of carbon. Results show that the
fractionation ratio, $^{12}$C/$^{13}$C, of the system varies with
radius and height in the disk. Different behaviour is seen in the
fractionation of different species. We compare our results with
$^{12}$C/$^{13}$C ratios in the Solar System comets, and find a stark
contrast, indicative of reprocessing.
\end{abstract}


\section{Introduction}

Studying the chemistry of the inner regions of protoplanetary disks
(PPDs) is a step into the unknown. Until recently observations of
molecules in PPDs could not probe the warmer, inner regions of the
disk where planets form. Detections of CO, HCO$^+$, H$_2$CO, C$_2$H,
CS, SiO, HNC, CN and HCN \citep{dut97,kas97,qi03,thi04,sem05} at
millimetre wavelengths can only tell us about the physics and
chemistry in the cold conditions at radii greater than $\sim$50\,AU
because of the limits of millimetre-wave sensitivity and spatial
resolution. We have to go to the infrared to investigate the hotter
material, and there comparatively few molecules have been
seen. Initially H$_2$, CO \citep{naj03,bri03,bla04} and H$_2$O
\citep{car04} were detected. Recently, the \textit{Spitzer Space
  Telescope} has added OH, C$_2$H$_2$, CO$_2$ and HCN to the tally
\citep{car08,lah06} and a few detections of very large polycyclic
aromatic hydrocarbon (PAH) molecules have been made \citep{gee06}. The
Keck Interferometer has also brought us detections of hot HCN and
C$_2$H$_2$ \citep{gib07}. The regions of the disk close to the star
are of immense interest because they are where terrestrial (and
larger) planets begin to form, as well as comets. Molecules in the
inner 30 or 40\,AU could be the building blocks of the kinds of
pre-biotic and biotic molecules we see in the Solar System today.

In the absence of observational evidence, we are left with theoretical
work. This is valuable in its own right for understanding the
processes in these regions, and for predicting observations for future
proposals and even future technologies, such as ALMA\footnote{The
Atacama Large Millimeter Array, due for completion in 2012
(\url{www.alma.info})}, which will be able to probe these hidden inner
disks. Previous modelling attempts of these very inner regions of PPDs
(R$<$10\,AU) have been few \citep{mar02,mil03,ilg04,ilg06,agu08}, but
have shown that these inner regions of PPDs are rich in molecules,
including some complex molecules, such as benzene \citep{woo07}.

In this paper we present chemical models of a protoplanetary T
Tauri-stage disk and a Minimum Mass Solar Nebula (MMSN), paying
particular attention to the fractionation of carbon-bearing
species. The fractionation in a particular molecule is defined as the
abundance of the $^{12}$C-bearing variant over the $^{13}$C-bearing
variant\footnote{When there is more than one $^{13}$C per molecule,
  the fractionation ratio is taken to be [$^{12}$C]/[$^{13}$C]}.
Fractionation in carbon gives us a method of labelling: identifying,
for instance, different regions of the disk and also identifying the
different evolutionary processes involved in the formation of the disk
-- each process leaves its isotopic signature. Thus initially we study
a typical protoplanetary disk, the physical aspects of which are
explained in Sect.~\ref{sec:modelphys}. We include a treatment of gas
heating and cooling, which is expanded upon fully in
Appendix~\ref{sec:heatcool}. In Sect.~\ref{sec:modelchem} we give a
description of the chemical network, and the various chemical and
photo-fractionation mechanisms including isotope-exchange
reactions. Finally, we present our findings on carbon fractionation in
protoplanetary disks, and set our results in both a Solar System and
interstellar context.

\section{The disk model: Physics}
\label{sec:modelphys}

In order to accurately model a PPD one has to take into account many
physical and chemical processes. One needs a thorough understanding of
disk hydrodynamics and magnetohydrodynamics, turbulence, accretion,
stellar spectrum, thermal balance, radiation transport, chemical
composition, dust composition, etc. Whilst there are efforts to
understand disks comprehensively and to model them accordingly with
the appropriate feedback and interactions, such models are
computationally very expensive and are very limited in their extent
\citep[e.g.,][]{tur07}. Thus simplifications are often made,
effectively decoupling different elements of the situation.

A major simplification is the separation of chemistry and physics when
solving the equations of hydrodynamics for the disk temperature and
density structure.
In order to make the problem tractable a static hydrodynamic disk
model is used to provide densities and dust temperatures throughout
the disk. From this the UV photon distribution can be calculated
\citep[e.g.,][]{van03}, assuming contributions to the UV field from
the stellar source and the interstellar medium. Few models have been
developed that determine the structure of the disk and the radiative
transfer self-consistently \citep[e.g.,][]{nom02,mil03}. The final
parameter required is that of the gas temperature. Some models
\citep[e.g.,][]{mar02} have assumed that the gas and dust temperatures
are identical but this significantly underestimates the gas
temperature in the disk surface \citep{kam04}. We calculate the gas
temperature in the disk from the density and dust temperature profiles
using a heating-cooling balance technique. This is a simpler approach
than a finding a self-consistent solution \citep[see][for
  example]{aik06}, but neglects the feedback interaction between
changes in the gas temperature and gas density. Once all the physical
parameters have been determined they can be used as inputs for the
chemical model.

In reality, protoplanetary disks are not static: there is the
large-scale motion of material accreting onto the star. There are also
convective motions, often parameterised into vertical mixing motions
and radial mixing motions. On all scales there is turbulence, which is
mostly likely driven by magnetic fields in the disk. Turbulence drives
mixing in both radial and vertical directions and models of mixing in
the inner \citep{ilg04} and outer disk \citep{wil06,sem06} have shown
this to be important. Material will also be gradually accreted
radially towards the central star. Here we ignore turbulent mixing but
mimic the accretion flow of material moving toward the star by moving
a number of parcels of gas from the outer edge of the model at 35\,AU
inward, along lines of constant scaleheight ($z_h$). These parcels are
spaced vertically in 0.1$z_h$ intervals above the midplane of the
disk, and we assume axisymmetry about the midplane.  Each parcel will
flow into the star in a time which can be derived from the principle
of mass conservation. The radial velocity is given by:
\begin{equation}
v_{\mathrm r} = \frac{\dot{M}}{2\pi\Sigma R} \qquad R\gg R_\star
\end{equation}
with $R$, the radius, $\dot{M}$, the accretion rate and $\Sigma$, the
surface density of the disk. The accretion timescale, $\Delta t$, is
approximately $\Delta R$/$v_{\mathrm r}$. Thus,
\begin{equation}
\Delta t \sim \frac{2\pi\Sigma R\Delta R}{\dot{M}}.
\label{eq:advectime}
\end{equation}
$\Sigma$ scales with 1/$R$ fairly closely for R$>$10\,AU
\citep{dal01,and07}, and so Eq.~\ref{eq:advectime} only has a small
dependence on $R$. A parcel starting at 35\,AU will pass into the star
in a time of approximately 0.41\,Myr. In our framework of seventy
radial points with a spacing of 0.5\,AU, each parcel will spend
$\approx$6\,000\,yr at each radial gridpoint before it is moved
inwards. This simplistic method has been treated in more detail, for
instance, in \citet[][eq. (10)]{aik99}.

\subsection{Disk structure}
\label{sec:structure}

To determine the physical structure of the disk we use a model kindly
supplied to us by Paola D'Alessio and based on \citet{dal99,dal01},
which models a flared disk using the $\alpha$-formulation
\citep{sha73}. The model has a central star with a temperature
($T_\star$) of 4\,000\,K, a mass ($M_\star$) of 0.7\,M$_\odot$, a
luminosity ($L_\star$) of 0.9\,L$_\odot$, and a radius ($R_\star$) of
2.5\,R$_\odot$ \citep[c.f.,][]{gul98}. The mass accretion rate
($\dot{M}$) is 10$^{-8}$\,M$_\odot$\,yr$^{-1}$. The viscosity
parameter, $\alpha$=0.01. The dust in the disk has an ISM size
distribution \citep{dra84} and is assumed to be well-mixed with the
gas. The dust disk extends from $\sim$0.1\,AU to 300\,AU and has a
mass of 0.032\,M$_\odot$. The model provides the density and grain
temperature. The gas temperature is calculated separately, and details
are given in the following section.

\begin{figure*}
\includegraphics[width=8cm]{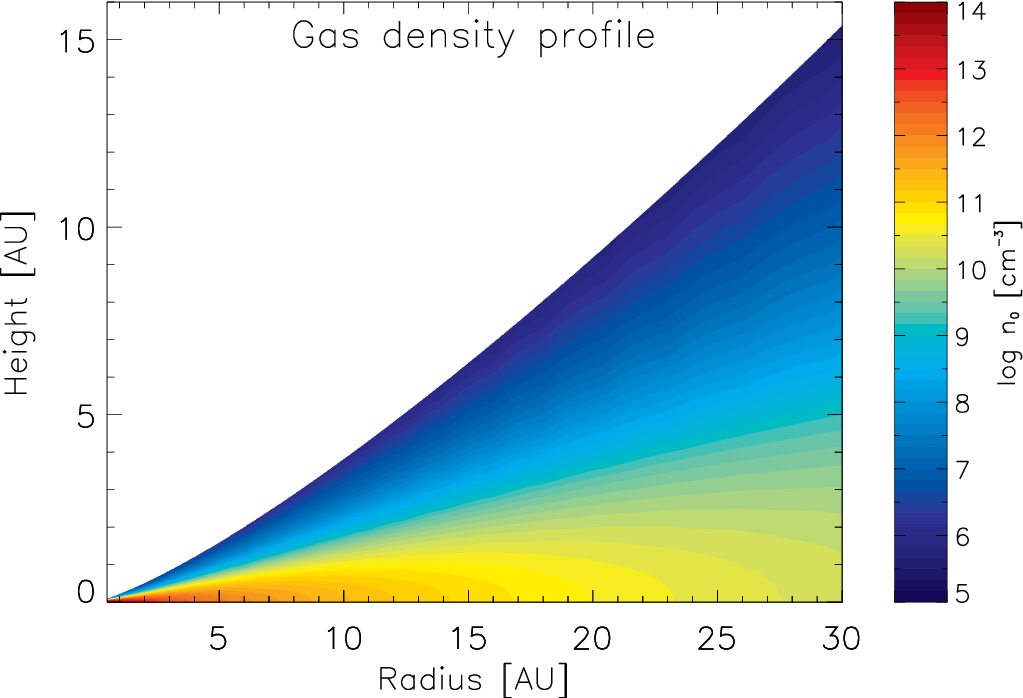}\includegraphics[width=8cm]{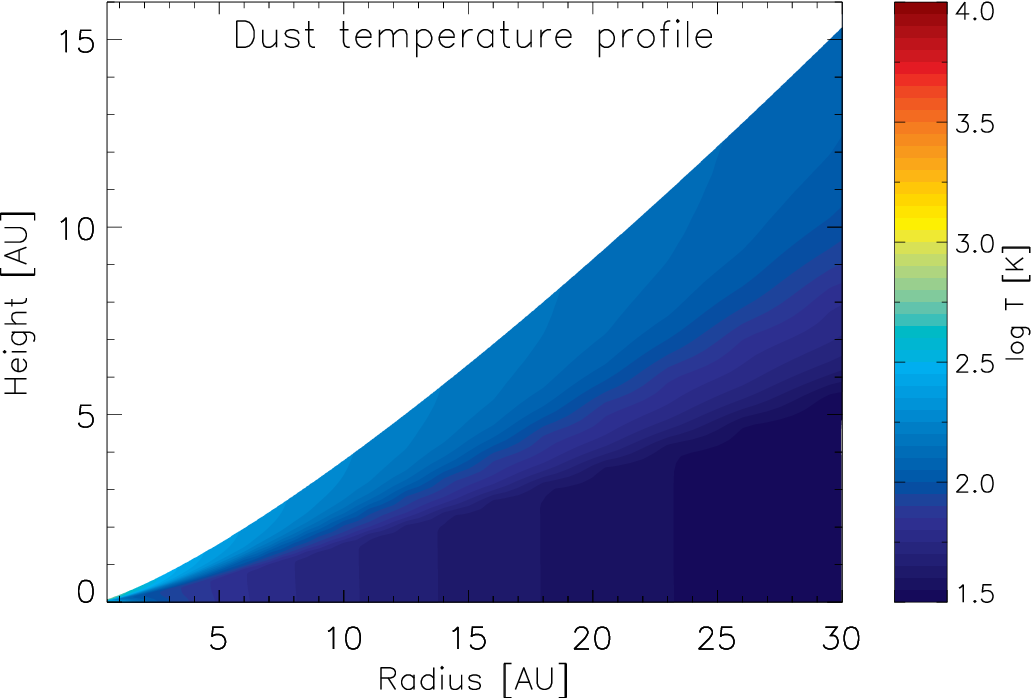}
\caption{Gas density and dust temperature profiles from the models of
Paola D'Alessio, re-gridded to our 35$\times$16\,AU grid, spaced at
0.5$\times$0.02\,AU. \label{fig:dustdisk}}
\end{figure*}

From this model we have calculated a gas pressure scaleheight ($z_h$)
for the disk based on the distance over which the density drops to the
0.01\% level. The radial dependence of $z_h$ can be fit by a power
law:
\begin{equation}
z_h = 0.0337R^{14/11}.
\end{equation}
Our chemical calculations extend from the midplane up to 6\,$z_h$.

\subsection{Gas and dust temperature}
\label{sec:gastemp}

\begin{figure}
\includegraphics[width=8cm]{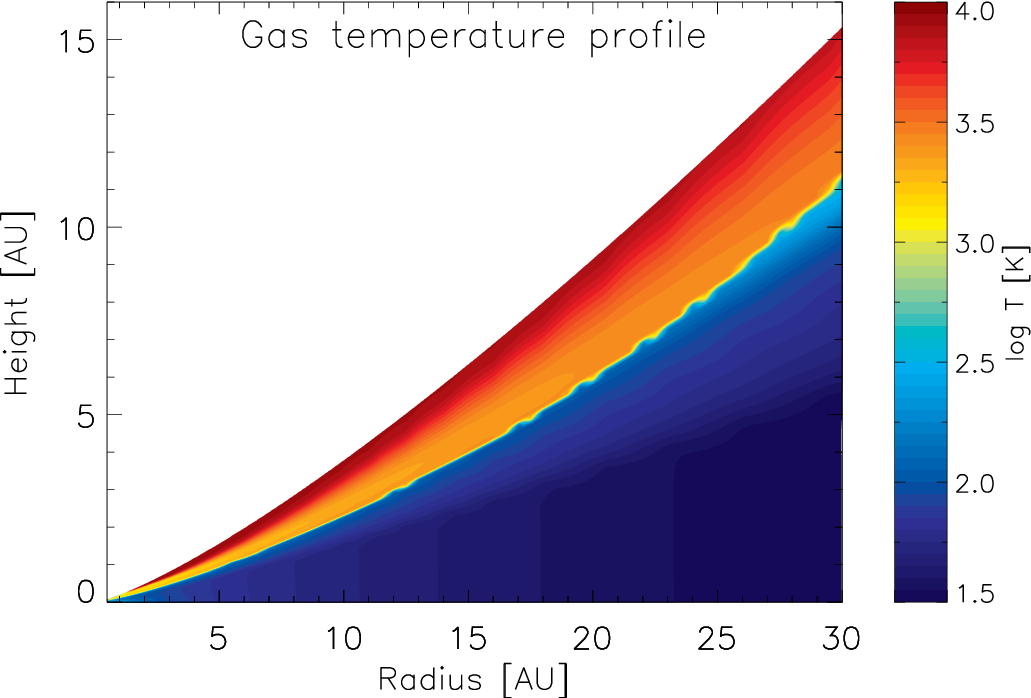}
\caption{Results of the gas temperature determination for the inner
 30\,AU of the disk model. The slight ``wobble'' in the transition
 from the cold disk interior to the hot surface layers is due to a
 change from bare to mantled grains, which affects the heating due to
 H$_2$ formation (Sect.~\ref{sec:gasheatcool}). \label{fig:gtemp}}
\end{figure}
\clearpage
\begin{deluxetable}{lll}
\tablecaption{\label{tab:heatcool} Gas heating and cooling mechanisms included.}
\tablehead{\colhead{Mechanism} & \colhead{Reference(s)} & \colhead{Notes}}
\startdata
\cutinhead{Heating}
Photoelectric effect & \citet{kam01} & For silicate grains \\
\nodata              & \citet{bak94} & For small graphite and PAH grains \\
H$_2$ collisional de-excitation & \citet{kam01} & \nodata \\
H$_2$ photodissociation & \citet{kam01} & \nodata \\
H$_2$ formation & \citet{kam04} & \nodata \\
\nodata         & \citet{caz02a,caz04} & H$_2$ formation efficiency\\
C ionisation    & \citet{tie85} & \nodata \\
Cosmic rays     & \citet{kam01} & For $\Sigma<$150g\,cm$^{-2}$ \\
\nodata         & \nodata       & \citep{ume81} \\
Stellar X-rays  & \citet{gor04} & X-ray ionisation rate \\
\nodata         & \citet{sha02} & Secondary effect \\
\cutinhead{Cooling}
O\textsc{i} fine structure lines& \citet{tie85} & \nodata    \\
\nodata                         & \citet{kam01} & Molecular line data \\
C\textsc{i} fine structure lines& \citet{tie85} & \nodata    \\
\nodata                         & \citet{kam01} & Molecular line data \\
C\textsc{ii} line at 157.7\,$\mu$m & \citet{tie85}   & \nodata    \\
\nodata                            & \citet{kam01}   & Molecular line data \\
$^1$D line of O\textsc{i} at 6300\,\AA & \citet{tie85} & \nodata    \\
\nodata                                & \citet{ste89} & Electron impact excitation \\
H$_2$ rovibrational lines & \citet{leb99} & \nodata    \\
CO rotational lines & \citet{tie85} & \nodata     \\
\nodata             & \citet{kam01} & Molecular line data \\
CH rotational lines & \citet{kam01} & \nodata\\
Lyman-$\alpha$ line & \citet{kam01} & \nodata\\ 
Gas-grain collisions& \citet{kam01} & \nodata
\enddata
\end{deluxetable}
\clearpage

Until fairly recently PPD models assumed that the gas portion of the
disk is in thermal equilibrium with the dust throughout its entirety
\citep[e.g.,][]{mar02} . This can often be an underestimate of the gas
temperature in regions of the disk close to the central radiation
source, and especially in the tenuous surface which is exposed to the
interstellar radiation field (ISRF), where conditions are similar to a
photon-dominated region (PDR). This assumption can underestimate the
gas temperature by an order of magnitude \citep{kam04}.

In an effort to more accurately model the isotope chemistry in PPDs,
for which gas temperature is an important factor (see
Sect.~\ref{sec:isotopes}), we have calculated the gas temperature
separately from the dust temperature, by balancing heating and cooling
terms. Several authors have previously taken this approach
\citep{kam01,kam04,gor04,tie85} and we follow the prescriptions given
in their papers, adapting them to T Tauri disks where necessary (see
Table~\ref{tab:heatcool} for details).

We take into account seven mechanisms which affect the gas heating
rate in the disk -- the photoelectric effect ($\Gamma_\mathrm{PE}$),
collisional de-excitation of H$_2$ ($\Gamma_\mathrm{CDx}$),
photodissociation of H$_2$ ($\Gamma_\mathrm{Phd}$), formation of H$_2$
($\Gamma_\mathrm{Form}$), ionisation of C ($\Gamma_\mathrm{CIon}$),
cosmic rays ($\Gamma_\mathrm{C-ray}$), and stellar X-rays
($\Gamma_\mathrm{X-ray}$). Gas-grain collisions can also act as a
heating mechanism when the dust temperature exceeds the gas
temperature. However, since this rarely happens in the disk model,
gas-grain collisions act mostly as a cooling mechanism
($\Lambda_\mathrm{G-G}$). Other cooling mechanisms included are: the
fine-structure lines of atomic oxygen at 63.2, 145.6 and 44.0\,$\mu$m
($\Lambda_{\mathrm{O}\textsc{i}}$), the fine-structure lines of atomic
carbon at 609.2, 229.9 and 369.0\,$\mu$m
($\Lambda_{\mathrm{C}\textsc{i}}$), the 157.7\,$\mu$m line of
singly-ionised carbon ($\Lambda_{\mathrm{C}\textsc{ii}}$), the
metastable $^1$D line of atomic oxygen at 6300\,\AA~
($\Lambda_\mathrm{O6300}$), rovibrational lines of H$_2$
($\Lambda_\mathrm{H_2}$), 25 rotational lines of CO
($\Lambda_\mathrm{CO}$), rotational cooling of the CH radical
($\Lambda_\mathrm{CH}$) and Lyman-$\alpha$ cooling
($\Lambda_\mathrm{Ly\alpha}$). Further details are given in
Appendix~\ref{sec:heatcool}.

Thus the temperature of the gas is calculated by balancing the heating
and cooling rates.  We use the modified Brent's method of root-finding
to solve for the gas temperature, as found in \citet{pre92}. This
method works well in regions where the gas temperature function is
continuous. Results of the gas temperature calculation are shown in
Fig.~\ref{fig:gtemp}.

To increase computational efficiency we only perform this calculation
once at every grid point, rather than iterating to a solution. Thus
the gas temperature is calculated first using chemical abundances from
the previous gridpoint and the new temperature used to calculate the
chemical reaction rates for the current grid point. This is acceptable
since there is only a weak coupling between the gas temperature
calculation and the chemistry: any changes in temperature in response
to a small change in chemistry will be small, and vice versa.


\subsection{Radiative transfer}

Protoplanetary disks are subject to radiation from the central stellar
source and from the ISRF. In order to calculate the UV radiation field
at any one point in our disk model, we make use of the direct/diffuse
treatment of FUV photons from \citet{ric00}. We include the effects of
radiation from both the central protostar and from the external UV
field, and also include the effects of scattering of photons by dust
grains. Absorption and scattering coefficients are based on the work
of \citet{dra84} and \citet{pre93}, who assume a mixture of silicates,
amorphous carbons and dirty ice-coated silicate grains. The disk is
divided up into square cells ($\Delta z$=$\Delta r$=0.02\,AU) and the
radiative transfer equation is solved separately for both sources of
radiation, and for the direct and diffuse (scattered) components of
each UV radiation field \citep[see][for details]{ric00}. The total UV
flux at a given point is the sum of the contributions from the stellar
and interstellar UV fields, that is, we employ a 1+1D approach rather
than the full 2D treatment calculated by \citet{ric00}.

The radiative transfer code is not able to account for the difference
between the shape of the interstellar UV field and that generated by
the spectrum of the young star. Observations have shown that T Tauri
stellar fields are different to the interstellar radiation field, and
can be dominated by strong emission features (such as
Lyman-$\alpha$). This can significantly impact the chemistry of the
surface layers of the disk since Lyman-$\alpha$ can dissociate some
molecules, such as OH and CH$_4$, and not others. \citet{ber03} found
that this effect can account for high CN/HCN ratios observed in some
disks \citep{dut97,kas97}. The strength of the T Tauri stellar
radiation field has been estimated from FUSE observations as a few
hundred times G$_0$ at a radius of 100\,AU \citep{ber03}, where G$_0$
is the standard interstellar radiation field strength. Here we adopt a
value of 500\,G$_0$ at a radius of 100\,AU.

We also calculate photoionisation by X-rays assuming an X-ray
luminosity of $10^{-4}$\,L$_\star$ \citep{gor04} or
$9\times10^{-5}$\,L$_\odot$.  The calculation of the X-ray
photoionisation rate is described in more detail in
Sec.~\ref{sec:chemistry} and the contribution of X-ray heating to the
gas temperature is discussed in Appendix~\ref{sec:heating}.

\subsection{Self-shielding}

Self-shielding by the abundant molecules H$_2$ and CO moderates the
effect of photodissociation. We incorporate this mechanism into our
model by using the \textquotedblleft shielding
factor\textquotedblright\ approach of \citet{van88} and
\citet{lee96}. For CO we use the data of \citet[][see
Sect.~\ref{sec:selfshield} for more detail]{van88}, and for H$_2$,
\citet{lee96}. These papers deal with interstellar clouds, for which
the turbulent line width can be larger than the thermal line widths in
PPDs. We use the re-normalisation technique proposed by \citet{a+h99a}
to account for this difference, where the column density of H$_2$ used
in the self-shielding calculation is reduced by the factor
$c_S$/3\,km\,s$^{-1}$ (where $c_S$ is the sound speed of the gas, and
3\,km\,s$^{-1}$ is the turbulent line width assumed in the cited
self-shielding calculations, for the ISM). In the inner disk, where
the temperature and the sound speed are highest, this has less effect
than that found by \citet{a+h99a} in the outer disk.

We follow \citet{a+h99a} in assuming that H$_2$ and CO are only
dissociated by the ISRF and not by stellar UV because of the
difficulty of solving the equation of radiative transfer
simultaneously in two dimensions. The stellar radiation is only
effective in the relatively low density surface regions (above
z/R$\sim$0.3, or 3.5\,$z_h$ at R=30\,AU), and thus has little effect
on the chemistry in the molecular regions we wish to study.

\section{The disk model: Chemistry}
\label{sec:modelchem}

\subsection{Isotopes of carbon}\label{sec:isotopes}

Our knowledge of the chemistry of $^{13}$C-bearing species stems from
30 years ago, and has advanced little since. This makes it difficult
to accurately study a system which includes $^{13}$C-bearing species
because reaction rates are simply not known. There is some data
available on differences in the zero-point energy between $^{12}$C-
and $^{13}$C-bearing species, but not for all species, and the Kinetic
Isotope Effect (KIE) has been used in theoretical work by some
\citep[e.g.,][]{you06} to account for the differences in molecular
mass, but this effect is complicated by the actual mechanism of
reaction, i.e., the KIE only applies when the bond being
broken/created is a bond between $^{13}$C and another atom. However,
there are two mechanisms which are known to fractionate carbon
isotopes -- fractionation through chemical exchanges, and through
photodissociation.

\subsubsection{Fractionation through chemical exchange reactions}

\citet{smi80} performed laboratory studies of isotope exchange
reactions, and revealed that the most important reaction for
exchanging carbon isotopes is:
\begin{equation}
\label{eq:C+iso}
\mathrm{^{13}C^+ + ^{12}CO \rightleftharpoons ^{13}CO + ^{12}C^+ +} \Delta E,
\end{equation}
as predicted by \citet{wat76}. $\Delta E$ is the zero-point energy
difference between the reactants and products, and is taken to be
35\,K for this reaction. This small energy difference makes the
forward reaction more efficient at low temperatures. The rate of this
reaction was initially measured by \citet{wat76} at 300\,K, and at 80,
200, 300 and 510\,K by \citet{smi80}. \citet{lan84} then calculated
rates down to 5\,K. A more recent calculation by \citet{loh98}
produced rates from 10-1\,000\,K, in reasonable agreement with both
\citet{smi80} and \citet{lan84}. The difference in the forward and
backward rates of Eq.~\ref{eq:C+iso} is small, but it is still
discernible even at 300\,K \citep{smi80}.

\citet{smi80} also measured the rate of another isotope-exchange
reaction at the same temperatures as before:
\begin{equation}
\label{eq:HCO+iso}
\mathrm{H^{12}CO^+ + ^{13}CO \rightleftharpoons H^{13}CO^+ + ^{12}CO +} \Delta E,
\end{equation}
with $\Delta E$=9\,K \citep{lan84}.
This is only effective at very low temperatures, with virtually no
difference in forward and backward rates at 300\,K and only a small
difference at 80\,K \citep{smi80}. \citet{lan84} calculated rates over
the range 5--300\,K and \citet{loh98} made rate calculations over
10--1\,000\,K.

Other exchange reactions similar to Eqs.~(\ref{eq:C+iso})
and~(\ref{eq:HCO+iso}) may occur. However, a chemical reaction will
almost certainly dominate over an exchange reaction such as reactions
(\ref{eq:C+iso}) or (\ref{eq:HCO+iso}) except when it is energetically
unfavourable to do so (due to complicated structural rearrangement,
for instance). In practice, many molecules react chemically with
C$^+$, for example, rather than undergo carbon exchange. An exception
to this is CS, for which $\Delta E\approx$26\,K in the reaction
\citep{wat76}. Another is CH$_3$ \citep{dal76}. \citet{loh98} suggests
that the reaction between HOC$^+$ and CO might be important for
isotope exchange since it moves in the opposite way to the exchange
involving HCO$^+$ (Eq.~\ref{eq:HCO+iso}), preferentially putting
$^{13}$C into CO. The difference in zero point energies is small,
2.5\,K. Similarly, \citet{lan92} suggested the importance of the
possible exchange between C$^+$ and CN, which might compete with the
photodestruction of CN in the upper layers of disks. The difference in
zero point energy is 34\,K. However, none of the rates of these
exchange reactions have been measured or calculated to our knowledge,
and we do not include them in our model.

For convenience, we have fit both the forward and backward reaction
rates for reactions (\ref{eq:C+iso}) and (\ref{eq:HCO+iso}) given in
the literature with a smooth function in the standard form. Thus for
reaction (\ref{eq:C+iso}),
\begin{eqnarray}
k_\mathrm{for} &=& 3.3\times10^{-10}\left(\frac{T}{300\,\mathrm{K}}\right)^{-0.448}~\qquad\mathrm{cm^3s^{-1}}\\
k_\mathrm{rev} &=& k_\mathrm{for}\exp(-35\,\mathrm{K}/T)\qquad\qquad\qquad\mathrm{cm^3s^{-1}}
\end{eqnarray}
and for reaction~(\ref{eq:HCO+iso}),
\begin{eqnarray}
k_\mathrm{for} &=& 2.6\times10^{-10}\left(\frac{T}{300\,\mathrm{K}}\right)^{-0.277}~\qquad\mathrm{cm^3s^{-1}} \label{eq:HCO+for}\\
k_\mathrm{rev} &=& k_\mathrm{for}\exp(-9\,\mathrm{K}/T)\qquad\qquad\qquad\mathrm{cm^3s^{-1}} \label{eq:HCO+rev}
\end{eqnarray}
These fits are in excellent agreement with both the experimental
results of \citet[][agree to within 7\%]{smi80} and the calculations
of \citet[][3\%]{lan84} and \citet[][8\% above 50\,K]{loh98}.

\subsubsection{Fractionation through photodissociation}
\label{sec:selfshield}

Carbon monoxide is dissociated mostly via line absorption and thus as
the column density of carbon monoxide towards the emitter increases,
so the absorption line saturates as photons are absorbed by the
intervening material. At some point, a carbon monoxide molecule
becomes shielded from dissociation by the presence of other molecules
along the line of sight -- not only molecules of the same kind, but
also isotopologues and atomic and molecular hydrogen, whose
dissociation bands may overlap. The degree of self-shielding depends
on this overlap, and thus the shielding of $^{13}$CO, say, will depend
on the column density of $^{12}$CO and H$_2$, as well as that of
$^{13}$CO. One must also take into account the attenuation provided by
dust grains.

This complex situation has been simplified by a number of authors for
different situations. \Citet{van88}\notetoeditor{This \LaTeX command
  is meant to produce a capitalised ``Van'', but it doesn't seem to
  work here. Maybe you have an alternative command.} treat $^{13}$CO
explicitly, and therefore we use their treatment in our model. They
have tabulated \textquotedblleft shielding
factors\textquotedblright\ which modify the photodissociation rate
according to the factors detailed above. These shielding factors
depend primarily on the column densities of $^{12}$CO and H$_2$, with
the other factors (including the isotope ratio) remaining fixed. Since
we calculate the column density of $^{13}$CO in our model (and this
differs from the fixed value of $N(^{12}$CO)/$N(^{13}$CO)=45 used in
\citealt{van88}), we assume $N(^{12}$CO)=45$\times N(^{13}$CO) for the
purpose of calculating the shielding factor for $^{13}$CO only. In
general, $N(^{12}$CO)/$N(^{13}$CO) is less than 45, and so we may be
slightly overestimating the contribution made to the self-shielding of
$^{13}$CO by $^{12}$CO.

For given column densities of $^{12}$CO and H$_2$, $^{13}$CO is
shielded less than $^{12}$CO, and thus is preferentially
photodissociated. So as one proceeds from a region of high extinction
to a region of low extinction, one would see $^{13}$CO being
photodissociated where $^{12}$CO is already self-shielded, and hence
an increase in the fractionation ratio, $^{12}$CO/$^{13}$CO, in this
region. This effect can be seen observationally at the edges of
molecular clouds \citep[e.g.,][]{she92}.

This situation is complicated further by some recent preliminary work
showing that self-shielding is temperature dependent \citep{lyo07}.
Since the predissociation bands for H$_2$ and CO are thermally
broadened and there is the possibility of overlap with adjacent
electronic states, additional CO vibrational bands have to be
considered at high temperatures. This work will be particularly
applicable to the temperatures found in PPDs, and we look forward to
the results of these investigations.

\subsubsection{Interstellar and Solar System context}

The isotope ratio for carbon ($^{12}$C/$^{13}$C) in the Solar System
is widely accepted to be 89 \citep{and89,cla04,mei07}, although recent
measurements of the solar photosphere have indicated a ratio of
80$\pm$1 \citep{ayr06}. This is greater than in the local interstellar
medium (ISM), where the value is taken to be 77 \citep{wil94}, greater
than the Orion Bar region
\citep[$^{12}$C/$^{13}$C$\sim$60;][]{kee98,lan90}, and much greater
than the Galactic Centre
\citep[$^{12}$C/$^{13}$C$\sim$20;][]{mil05,lan84}. This galactic
gradient \citep{lan90} is due to the higher star formation rate in the
inner Galaxy \citep{tos82}, where the fraction of $^{13}$C has been
enhanced by the $^{13}$C-rich ejecta of evolved, intermediate-mass
stars \citep{ibe83} in the time since the formation of the Solar
System.

\citet{wil94} give a numerical evaluation of how the isotope ratio
changes with galactocentric distance. Since the Sun formed
approximately 1.9\,kpc closer to the Galactic Centre \citep{wie96}
than its present location \citep[7.94\,kpc,][]{eis03}, presumably it
would have formed in a region with a lower $^{12}$C/$^{13}$C ratio,
viz., $\approx$67, ignoring temporal evolution. \citet{wie97} discuss
a method of incorporating the temporal evolution of the ISM, and
derive a value of 62 for the region and time period in which the Solar
System condensed. The large difference between this value and the
present Solar System value of 89 indicates that the Solar System must
have become either significantly enriched in $^{12}$C, or there must
have been a significant depletion of $^{13}$C. Recent studies of iron
isotopes in the Solar System have indicated that the Sun most likely
formed close to one or more massive stars \citep{hes04}, which
produce $^{12}$C in the triple-$\alpha$ reaction in their interiors
\citep{tim95}. These stars, which go on to form Type II supernovae,
may have contaminated the solar protoplanetary nebula with
$^{12}$C-rich material during its formation. Alternative explanations
for this difference in isotope ratios are X-ray flares \citep{fei02},
cloud mergers, orbital diffusion or radial gas streaming
\citep[see][for further details]{mil05}. It seems unlikely that the
increase in the $^{12}$C/$^{13}$C ratio above the interstellar value
is due to the processing of material once the Solar System had been
established \citep{lec98}.

In light of these factors, we choose an initial $^{12}$C/$^{13}$C
ratio of 77, between the values of 62 and 89, in line with that of the
present-day ISM ratio. The value of this ratio is not crucial to the
chemistry, and similar fractionation levels are obtained for the
entire range of values, 62--89.

\subsection{The reaction set}
\label{sec:chemistry}

\begin{deluxetable}{llccccccc}
\tablecaption{\label{tab:initabunds} Initial fractional abundances for
abundant carbon-bearing species. These abundances are the result of
1\,Myr in a molecular cloud model}
\tablewidth{0pt}
\tablehead{
\colhead{Species} && \colhead{Gas phase} && \colhead{$^{12}$CX/$^{13}$CX} && \colhead{Solid phase} &&  \colhead{g--$^{12}$CX/g--$^{13}$CX}}
\startdata
C$_2$      && 1.7\,(-7) && 169 && negligible&& --\\
C$_3$H$_2$ && 4.1\,(-7) && 160 && 1.9\,(-6) && 133\\
C$_2$H$_2$ && 3.6\,(-7) && 153 && 1.2\,(-6) && 141\\
C$_4$H     && 5.3\,(-7) && 152 && negligible&& --\\
HCN        && 2.4\,(-8) && 152 && 3.7\,(-6) && 111\\
HC$_3$N    && 1.1\,(-7) && 150 && 5.8\,(-7) && 134\\
CN         && 1.3\,(-7) && 146 && negligible&& --\\
HNC        && 1.0\,(-8) && 144 && 2.6\,(-7) && 106\\
C$_4$      && 7.2\,(-8) && 144 && negligible&& --\\
C$_3$      && 9.8\,(-8) && 134 && negligible&& --\\
C$_3$H     && 5.3\,(-7) && 133 && negligible&& --\\
C$_4$H$_2$ && 9.6\,(-9) && 133 && 8.5\,(-7) && 136\\
C$_2$H     && 6.5\,(-8) && 118 && negligible&& --\\
CH         && 2.4\,(-8) && 113 && negligible&& --\\
CH$_3$     && 4.8\,(-8) && 107 && negligible&& --\\
H$_2$CO    && 4.7\,(-8) && 107 && 2.1\,(-7) && 106\\
CH$_4$     && 2.3\,(-6) && 102 && 9.2\,(-6) && 99\\
C          && 1.1\,(-7) && 93  && negligible&& --\\
CO         && 3.1\,(-5) && 55  && 2.6\,(-6) && 55\\
CO$_2$     && 1.7\,(-7) && 51  && 6.4\,(-7) && 55
\enddata

\tablecomments{$x$\,($-y$) here represents $x\times10^{-y}$. Isotope
ratios for species with more than one carbon atom are calculated by
dividing the total abundance of $^{12}$C by the total abundance of
$^{13}$C in that particular molecule. Other species of note, and their
initial fractional abundances: H$_2$ 5.0$\times$10$^{-1}$, He
1.4$\times$10$^{-1}$, g--H$_2$O 1.4$\times$10$^{-4}$, H
2.5$\times$10$^{-5}$, g--NH$_3$ 1.6$\times$10$^{-5}$, O
1.4$\times$10$^{-6}$, OH 6.4$\times$10$^{-7}$.}
\end{deluxetable}

The chemical reaction network is based on the UMIST99 gas-phase
ratefile \citep{let00}, with two additions: 

1) we include grain surface reactions from \citet{has93} and
\citet{has92}. We consider cosmic ray heating and thermal desorption,
with updated binding energies from \citet{bis06}, \citet{obe05} and
\citet{fra01}. For atomic H, we use updated binding energies from
\citet{caz04,caz02a}, who treat physisorption and chemisorption onto
grains distinctly, with binding energies $E_\mathrm{H_p}$=600\,K and
$E_\mathrm{H_c}$=10\,000\,K, respectively. In our calculations we
assume that the sticking coefficient of H is 0.4, and that for all
other species is 0.3.

2) we include a simple treatment of X-ray chemistry similar to that by
   \citet{gor04}, i.e., we calculate an ionisation rate per atom,
   which depends on photon flux and cross-section:
\begin{equation}
\zeta_\mathrm{X}^i =
6.25\times10^8\int_{0.5}^{10}\sigma_\mathrm{X}^i(E)\frac{F(E)}{E}\exp(-\tau_\mathrm{X}(E))dE\quad\mathrm{s^{-1}},
\end{equation}
for atom $i$. Fits to X-ray cross-sections ($\sigma_\mathrm{X}^i$) for
   astrophysically relevant molecules are taken from
   \citet{ver95}\footnote{Please contact the authors for our fitting
   coefficients}. $F(E)$ is the X-ray photon flux at a radius R, as a
   function of energy
   0.5$<E<$10\,keV. $\tau_\mathrm{X}(E)=N_0\sigma_\mathrm{X}(E)$ is
   the X-ray extinction due to a column density $N_0$. We assume that
   X-ray ionisation leads to the loss of a single election, and that
   the ionisation rate for molecules is the sum of the rates for the
   constituent atoms. See \citet{aik01} for a more detailed treatment
   of X-ray chemistry.

In total, our reaction network comprises 475 gas and grain species,
and over 8\,000 gas-phase and surface reactions. More than
three-quarters of these reactions involve $^{13}$C. There have been
recent observations \citep{sak07,tak98} of $^{13}$C isotopologues and
isotopomers which show that there may be differing properties for
carbon atoms attached in different places in the molecule. However,
due to a dearth of experimental data on how $^{13}$C-bearing species
react in comparison to $^{12}$C-bearing species, we have assumed that
reactions involving $^{13}$C proceed at the same rate as their
$^{12}$C counterparts. We have taken account of the increased number
of reaction products which comes from the inclusion of isotopic
species. For example, the reaction:
\begin{equation}
\rm C_2 + H \longrightarrow\rm CH + C
\end{equation}
with rate $k$ now becomes:
\begin{eqnarray}
\rm^{12}C^{12}C + H &\longrightarrow&\rm ^{12}CH + ^{12}C\\
\rm^{12}C^{13}C + H &\longrightarrow&\rm ^{12}CH + ^{13}C\\
\rm^{12}C^{13}C + H &\longrightarrow&\rm ^{13}CH + ^{12}C\\
\rm^{13}C^{13}C + H &\longrightarrow&\rm ^{13}CH + ^{13}C,
\end{eqnarray}
with rates $k$, $k$/2, $k$/2 and $k$, respectively. We preserve
functional groups such that, e.g.,
\begin{eqnarray}
\rm ^{13}CH_3^{12}CN + He^+ &\longrightarrow&\rm ^{12}CN + ^{13}CH_3^+
+ He\\ \rm ^{13}CH_3^{12}CN + He^+ &\crossarrow&\rm ^{13}CN
+ ^{12}CH_3^+ + He
\end{eqnarray}
and we preserve double bonds in preference to single bonds:
\begin{eqnarray}
\rm H_2^{13}CO + ^{12}CH  &\longrightarrow&\rm H^{13}CO + ^{12}CH_2\\
\rm H_2^{13}CO + ^{12}CH &\crossarrow&\rm H^{12}CO + ^{13}CH_2.
\end{eqnarray}

As initial fractional abundances we use the outputs of an interstellar
(IS) cloud model ($n$=2$\times$10$^4$\,cm$^{-3}$, $T$=10\,K,
$A_\mathrm{V}$=10) which uses the same chemical network, and is
allowed to run for 10$^6$\,yr.  The inputs to this cloud model are the
\textquotedblleft low metal abundances \textquotedblright of
\citet{gra82}, viz. H:He:O:C:N:Si are
1:0.14:1.76$\times$10$^{-4}$:7.30$\times$10$^{-5}$:2.14$\times$10$^{-5}$:2.00$\times$10$^{-8}$.
Our IS cloud model is a simple single-point approximation, but it
reproduces the results of \citet{lan89} very well. We are in agreement
to within a factor of 2 for important species such as CO, HCO$^+$, O,
CH, CH$_2$ and H$_2$CO, and a factor of five agreement with C$^+$,
C$_2$ and H$_2$O. There is a lesser agreement with CN, HCN, C$_2$H,
CH$_4$ and OH due to the advances in the accuracy of reaction rate
determination in the last 25 years. For instance, our model produces
an overabundance \citep[compared to][]{lan89} of CN by a factor of
$\sim$20, due to a faster rate for the reaction:
\begin{equation}
\rm CH + N \longrightarrow\rm CN + H.
\end{equation}
\citet{lan89} use a rate of 4.5$\times$10$^{-12}$\,cm$^3$\,s$^{-1}$ at
20\,K, whereas the revised rate from \citet{let00} is
2.1$\times$10$^{-10}$\,cm$^3$\,s$^{-1}$, 47 times faster. See
Table~\ref{tab:initabunds} for input abundances for select species,
and also the fractionation ratios after the molecular cloud stage.

 The chemical fractionation which occurs in an interstellar cloud is
explored in detail by \citet{lan84} and \citet{lan89}. In summary,
\citet{lan84} are able to classify C-bearing species into three
families -- CO, HCO$^+$ and the ``carbon isotope pool'' -- with
distinct isotopic behaviours. The fractionation in CO is driven mainly
by reaction~(\ref{eq:C+iso}), which favours the production of
$^{13}$CO at the low temperatures and densities of interstellar
clouds, driving the $^{12}$CO/$^{13}$CO ratio down. The fractionation
of HCO$^+$ is driven both by reaction~(\ref{eq:HCO+iso}), which
preferentially puts $^{13}$C into HCO$^+$ at low temperatures, and
also by its formation from elements of the carbon isotope pool, which
are $^{12}$C-enriched. The fractionation in these remaining
carbon-bearing molecules is driven to high $^{12}$C/$^{13}$C ratios
since the chemistry is based on C$^+$, for which $^{12}$C$^+$ is
favoured at low temperatures (reaction~\ref{eq:C+iso}) and much
$^{13}$C is taken up in $^{13}$CO.


\section{Results}
\label{sec:results}

\subsection{Gas heating and cooling}
\label{sec:gasheatcool}

The gas temperature in our model of a circumstellar disk can reach
$\sim$8\,000\,K at the surface, vastly exceeding the temperature of
the dust from the input dust models, which is a few hundred degrees in
the same region. Such an effect has also been found by
\citet{kam01,kam04} and \citet{gla04} using different disk
models. Clearly there is a strong need for disk models which calculate
dust and gas temperatures individually. The gas and dust are not well
coupled collisionally (and hence not in thermal equilibrium) in the
disk surface, where the optical depth to the ISRF ($\tau$) is less
than 0.5 ($\sim$3\,$z_h$).

Heating by the photoelectric effect on small carbon and PAH grains is
very important in the surface layers of the disk. This mechanism
dominates almost all others in the upper third of the disk; gas
heating due to H$_2$ formation becomes the most effective mechanism in
a limited region where H$_2$ is photodissociated, around
$z$=4--5$z_h$.  Intermediate layers of the disk are heated mainly by
X-ray heating for radii $\gtrsim$10\,AU, and by gas-grain collisions
inside of this radius. The midplane is heated predominantly by
collisional de-excitation of H$_2$ molecules. The lower two-thirds of
the disk are cooled mostly by molecules - in the most part, CO
molecules are the most effective coolant, although CH molecules cool
the midplane very efficiently at larger radii, R$\gtrsim$17\,AU. In
the upper third of the disk, cooling by the forbidden lines of O{\sc
i} is very effective, with Lyman-$\alpha$ cooling dominating at the
very surface of the disk, inside of 20\,AU.

The transition between the hot surface layers and the cool bulk of the
disk shows some small oscillations in temperature
(Fig.~\ref{fig:gtemp}). This is a result of the inclusion of icy
grains, with a different H atom binding energy compared to bare grains
in our model. On icy grains H atoms can only physisorb and therefore
have a relatively short residence time. The oscillations occur in the
region where the transition from bare to icy grains occur.


\subsection{Fractionation of carbon isotopes}

The disk can can be split into different regions depending on
temperature, which is the dominating factor in the carbon
fractionation of different species. The cold midplane region (``the
cold region'', $T$=30--100\,K) of the disk extends up to
$\sim$3.5\,$z_h$, which equates to 9\,AU at a radius of 30\,AU. In
this region the majority of molecules are frozen out onto the surface
of dust grains; only very volatile species (CO, CH$_4$) are in the gas
phase. The warmer temperatures in the very inner part of the disk
means that the ice line for water falls close to 2\,AU.

Above the cold region is a transition region (``the transition
region'', $T$=100--2\,500\,K), which generally has a thickness of
$\sim$0.8\,$z_h$ (2\,AU at $R$=30\,AU). The increase in temperature in
this region causes molecules to evaporate from grain surfaces - H$_2$O
is one of the last species to persist on grains. UV extinction is low
enough for most molecules to be photoprocessed, although H$_2$,
$^{12}$CO and $^{13}$CO remain shielded.

Lastly, the surface region (above $\sim$4.3\,$z_h$) of the disk is
heavily ionised and very hot (``the hot region'',
$T$=2\,500--8\,000\,K). Ionised species have a fractional abundance of
$\sim$10$^{-3}$ in the surface region, indicating that most hydrogen
is atomic and not yet ionised. These three regions are evident in
Figs.~\ref{fig:COfrac} and \ref{fig:C+frac}. We will look at some
characteristic species in detail and their properties in these three
regions.

\subsubsection{CO, HCO$^+$ and CO$_2$}
\begin{figure*}
\includegraphics[width=16cm]{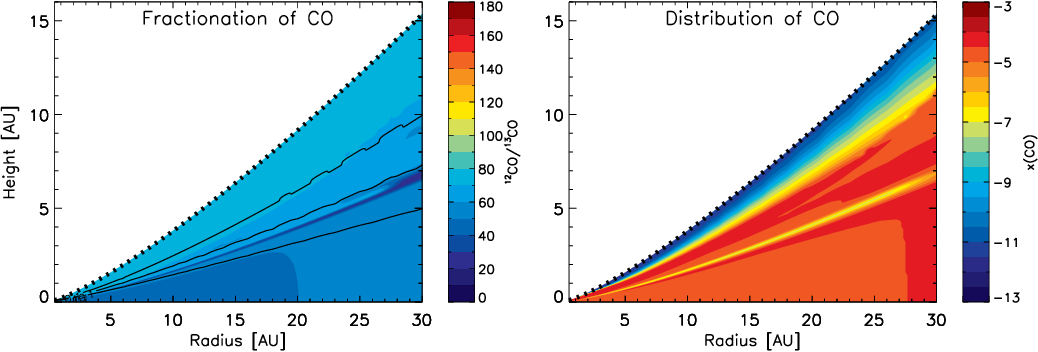}
\caption{Carbon monoxide fractionation shown as the ratio of
  $^{12}$CO/$^{13}$CO throughout the disk. Three different regions can
  be seen in the fractionation data - the hot region at the top of the
  disk, the cold region at the bottom and the transition region in
  between. The solid lines in this and following figures represent
  levels of optical depth to the ISRF, $\tau$=0.1, 1, 10. The dotted
  line indicates the designated surface of the disk at 6\,$z_h$.}
\label{fig:COfrac}
\end{figure*}

CO is an important species in disks since it is the dominant gas-phase
molecule other than H$_2$ and can be used observationally to trace the
bulk gas of the disk. It is also important because it can be used to
trace the vertical temperature structure in disks, as in
\citet{gui98,dar03} and \citet{pie07}.

Of the CO introduced into the disk from the interstellar medium at
35\,AU in our model, 92\% of that is in the gas phase, and 8\% is in
the form of CO frozen onto dust grains. CO desorbs at low temperatures
\citep[26\,K;][]{obe05} and in the inner disk temperatures are always
above this, so CO quickly desorbs and very little ($\ll$1\%) CO ice
remains. Hence there is a high abundance of gaseous CO available to
take part in chemical reactions.

\begin{figure}
\includegraphics[width=16cm]{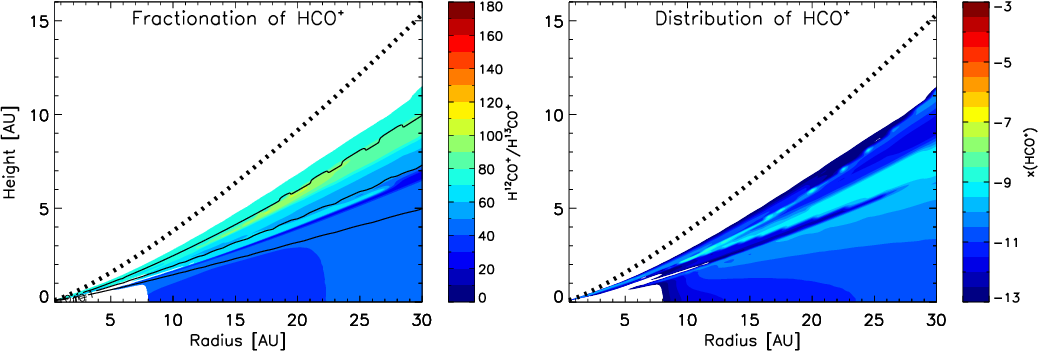}
\caption{The fractionation and distribution of the formyl ion,
  HCO$^+$. White areas in this and subsequent figures indicate regions
  of negligible fractional abundance, $x$(X)$<$10$^{-13}$.}
\label{fig:HCO+dist}
\end{figure}

The fractionation ratio of CO in the disk has a limited range, varying
from 25-77 throughout the disk. This contrasts with observed ratios in
diffuse interstellar clouds where there is a much wider range,
15$<$N($^{12}$CO)/N($^{13}$CO)$<$170 \citep{lis07}.  The fractionation
ratio of CO in the midplane does not change a great deal, 47--55
between 1 and 30\,AU (Fig.~\ref{fig:COfrac}). However its high
abundance means it is able to be involved in exchange reactions
thereby altering the fractionation of other molecules, e.g., HCO$^+$,
which shows an increase in fractionation with decreasing radius. This
is a consequence of exchange reaction~(\ref{eq:HCO+iso}). Under the
high density conditions of the disk, this process is roughly in
chemical equilibrium:
\begin{eqnarray}
k_\mathrm{for}n(\mathrm{H^{12}CO^+})n(\mathrm{^{13}CO}) &=& 
           k_\mathrm{rev}n(\mathrm{H^{13}CO^+})n(\mathrm{^{12}CO})\\
\Rightarrow \frac{n(\mathrm{H^{12}CO^+})}{n(\mathrm{H^{13}CO^+})} &=& 
         \exp{(-9\,\mathrm{K}/T)}\frac{n(\mathrm{^{12}CO})}{n(\mathrm{^{13}CO})}
\end{eqnarray}
using the relation in Eq.~(\ref{eq:HCO+rev}). Thus at temperatures of
32\,K, where the $^{12}$CO/$^{13}$CO ratio is 55, we should expect
H$^{12}$CO$^+$/H$^{13}$CO$^+$=41, and at temperatures of 270\,K, 45,
as seen. Thus at higher temperatures (T$\gtrsim$60\,K), the
fractionation of HCO$^+$ begins to trace the fractionation of CO, at
least for the midplane region of the disk where CO is the only source
of HCO$^+$. The fractionation of HCO$^+$ is shown in
Fig.~\ref{fig:HCO+dist}. Within 10\,AU, HCO$^+$ reacts with
increasingly abundant hydrocarbons (e.g., C$_3$H$_4$, C$_4$H$_2$) and
ammonia to create reactive ions and to reform CO, and within 8\,AU
these reactions become faster than the interchange between CO and
HCO$^+$.

At radii of a few AU, the reaction between CO and OH is the major
contributor to the formation of CO$_2$. This reaction has a moderate
activation barrier, and thus proceeds more rapidly at higher
temperatures. This reaction, which is important for both atmospheric
and combustion chemistry, is one of the few for which mass-independent
isotope effects have been investigated
\citep[c.f.,][]{che05,ste80,smi82,roc98}. These calculations and
experiments find a small difference in the reaction rates of $^{12}$CO
and $^{13}$CO with OH, but they were carried out at significantly
higher pressures than are found in protoplanetary disks. At the lowest
pressures considered, they found that the reaction involving $^{13}$CO
is faster than that involving $^{12}$CO by less than 1\%. Given that
this percentage is dwarfed by the uncertainties in published reaction
rates, the Kinetic Isotope Effect (of which this is an example) is
unlikely to affect our results.

The fractionation ratio of CO ice in the midplane resembles that of
gaseous CO. CO$_2$ ice, with a higher binding energy than CO ice, has
a fractionation ratio which decreases from its input value of 55 to a
value slightly higher than that of CO (46) in the very inner part of
the disk, 49. The value of $^{12}$CO$_2$/$^{13}$CO$_2$ ice ratio has
been determined to be 81$\pm$11 in a young protostar, \object{Elias
  29} \citep{boo00a}. This is unusually high, especially since the
nascent cloud, $\rho$ Oph, shows $^{12}$C/$^{13}$C ratios typical of
the local ISM \citep{cas05,ben01}. This value is also somewhat higher
than the $^{12}$CO/$^{13}$CO ice ratio along the same line of sight,
71$\pm$15 \citep{boo02}. A suggestion for this high value
\citep{boo00b} is that CO$_2$ may have been formed from C$^{(+)}$
rather than CO, as generally assumed. In our model, the vast majority
of CO$_2$ ice comes from CO, in agreement with \citet{ehr00}. Whatever
the formation route, it seems that CO$_2$ ice ratios are determined in
the parent cloud rather than the hot core around the protostar
\citep{cha04}.

Towards the top of the cold region in the disk ($z\sim$2.7\,$z_h$)
there is a layer of low abundance of CO. Above this layer, CO can be
destroyed by reactions with He$^+$, from the X-ray ionisation of
He. Some carbon ions resulting from this destruction end up forming
hydrocarbons on the surfaces of dust grains. At this height in the
disk, the grain temperature is just high enough that hydrocarbons can
thermally desorb, ensuring that the carbon contained in them is not
lost from the gas. In contrast, in the low abundance layer at
2.7\,$z_h$, hydrocarbons that form on the grains are retained,
resulting in a loss of carbon from the gas, made evident in a loss of
CO. Below this low abundance layer, X-rays do not penetrate and the
destruction rate of CO by reactions with He$^+$ or H$_3^+$ is
significantly lower. This effect is greater in the MMSN model
(Sect.~\ref{fig:MMSN}) because that model has a higher density and a
lower temperature, i.e., more efficient freezeout and less efficient
desorption for a given column density.

In the transition region of the disk, fractionation is driven strongly
by photoprocesses. HCO$^+$ becomes abundant, with fractional
abundances of 1--10$\times$10$^{-10}$, and consequently increases the
fractionation ratio of CO, through reaction (\ref{eq:HCO+iso}).  Due
to self-shielding effects, CO is more resilient to photodissociation
than other carbon-bearing species and thus is molecular in a region
where other molecules are starting to become dissociated by UV
photons. There is also a difference in the degree of self-shielding
between $^{12}$CO and $^{13}$CO, and thus there is a narrow layer in
which $^{13}$CO is photodissociated and $^{12}$CO is not. This
``photo-fractionation'' layer occurs at the very top of the transition
region ($z\sim4.1\,z_h$), and is minimal in thickness due to the low
column densities to the dissociating UV radiation fields at the top of
the disk. This region of the disk is also the one in which the
exchange between CO and C$^+$ (reaction \ref{eq:C+iso}) is most
important. C$^+$ and $^{13}$C$^+$ are abundant because CO and
$^{13}$CO are being photodestroyed, and isotope exchange and
photodissociation compete.

\subsubsection{C$^+$}

\clearpage

\begin{figure}
\includegraphics[width=7.5cm]{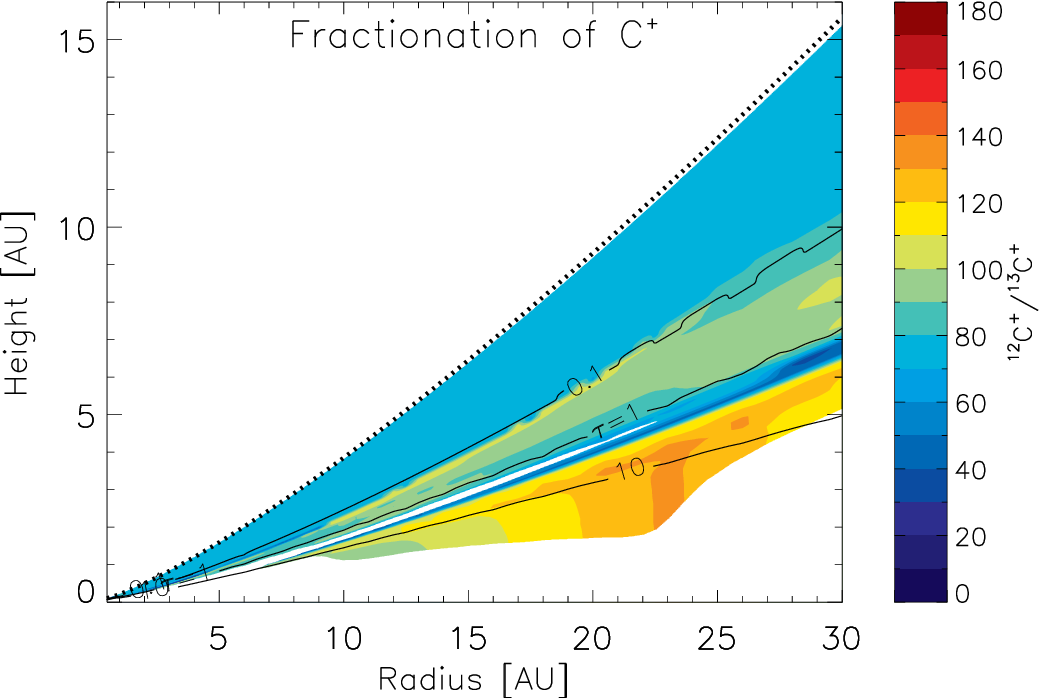}
\includegraphics[width=7.5cm]{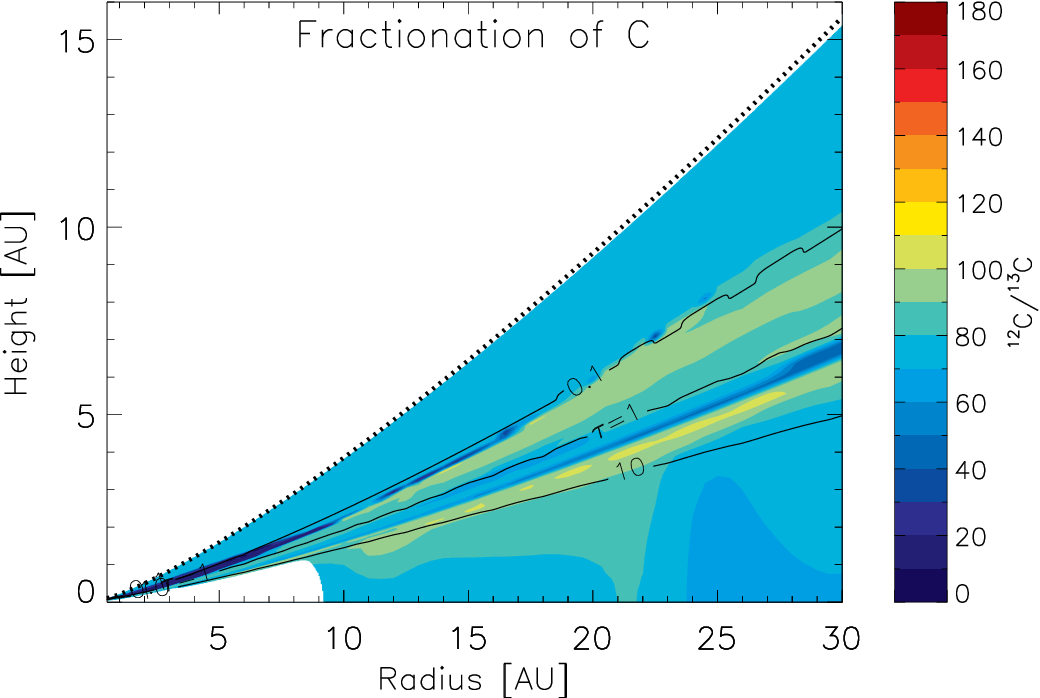}
\caption{Variation of $^{12}$C$^+$/$^{13}$C$^+$ and $^{12}$C/$^{13}$C
throughout the disk.}
\label{fig:C+frac}
\end{figure}

\clearpage

Fractionation in the top layer of the disk is relatively easy to
understand, since all but trace amounts of carbon are to be found in
the form of C$^+$ (and $^{13}$C$^+$). Thus it follows that the
fractionation ratio in this region must reflect the ``input''
value. For our fiducial model, this input value is 77, and
Fig.~\ref{fig:C+frac} shows this value in the upper layer of the
disk. C$^+$ is produced very rapidly, on a timescale of days, mainly
by photoionisation of C. This is much faster than a vertical or radial
mixing timescale, and thus the fractionation in the upper ionised
region is likely to persist as a long-lived feature of the disk. This
provides us with a simple way to quantify the carbon fractionation in
an observed disk.

The fractionation pattern in the upper, hot layer of the disk shows
the need for gas temperature calculations in modelling. Assuming that
gas and dust temperatures are identical produces a very uniform disk,
with slight increases in fractionation in the transition region, where
the difference in self-shielding factors for carbon monoxide is
evident. The upper region of the disk has the same degree of
fractionation as lower levels, and in general, carbon-bearing species
do not reflect the input fractionation ratio. In our model, where the
gas temperature is calculated separately from the dust temperature,
the activation barrier of reaction~(\ref{eq:C+iso}) becomes negligible
at sufficiently high temperatures, such as those found in the disk
surface layers. Thus the fractionations in C$^+$ and CO are averaged
into the input fractionation ratio of 77.

\subsubsection{H$_2$CO, C and the carbon isotope pool}

\begin{figure*}
\includegraphics[width=16cm]{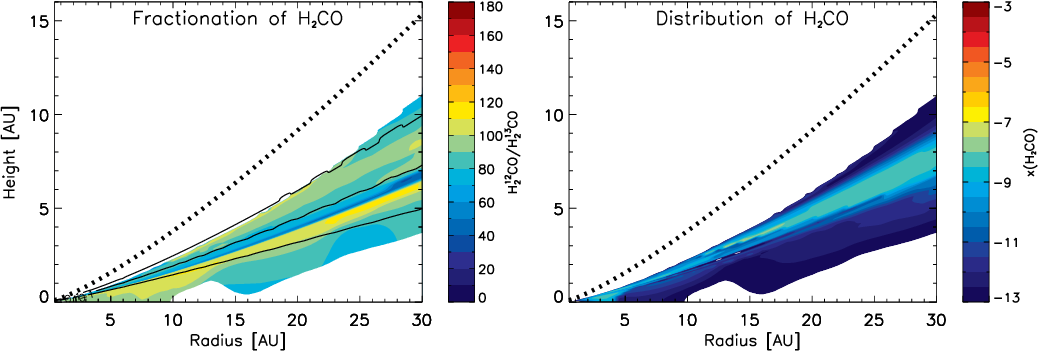}
\caption{Formaldehyde fractionation shown as the ratio of
  H$_2^{12}$CO/H$_2^{13}$CO throughout the disk. }
\label{fig:H2COduo}
\end{figure*}

The fractionation in H$_2$CO in interstellar clouds provides one of
the upper bounds to the total $^{12}$C/$^{13}$C ratio
\citep{lan84}. However, this is not the case in disks, where the
fractionation in H$_2$CO varies from $\sim$60-100
(Fig.\ref{fig:H2COduo}). H$_2$CO is formed almost exclusively by the
reaction between CH$_3$ and O in the gas phase, and thus the
fractionation in H$_2$CO reflects that in atomic carbon.

The fractionation ratio of atomic carbon in the disk varies from
16--110 (see Fig~\ref{fig:C+frac}). In general this can be further
constrained to $\sim$45--110, with a thin layer 3.3--4.5\,$z_h$
(0.5$<$R$<$10\,AU) where the ratio drops precipitously to
$\sim$15. This drop is due to a slight enhancement of the
photodissociation rate of $^{13}$CO over $^{12}$CO due to the
differences in self-shielding, causing the abundance of $^{13}$C to
rise. Atomic carbon is the basis for the formation of many
hydrocarbons, and thus species in the carbon isotope pool (e.g., CH,
CH$_4$, etc.) will follow the fractionation in C.


\subsubsection{Nitrogen-bearing species}

Solid HCN is a major repository of carbon in the cold regions of the
disk, storing up to 5\% of all available carbon, as well as $\sim$20\%
of all nitrogen. Desorption of HCN from grain surfaces becomes
efficient at a height of $\sim$3.2--3.4$z_h$.  HCN is then destroyed
in the gas phase by photons. What happens to the liberated nitrogen
depends on position in the disk. At 3.5$z_h$ it is cycled back into
HCN through reactions with carbon clusters (C$_n$). At 3.7$z_h$, the
availability of oxygen-bearing reactants derived from thermally
desorbed H$_2$O is much greater than at 3.5$z_h$, and so N is cycled
into NO, and C is cycled into CO through reactions with OH. This
causes a slight increase in the abundance of CO, which can be seen in
the right panel of Fig.~\ref{fig:COfrac}.

Solid HCN retains the fractionation ratio of the interstellar cloud
model, H$^{12}$CN/H$^{13}$CN = 111. In the gas phase, in intermediate
layers of the disk, the fractionation of HCN, HNC and CN is controlled
by, and mimics, the fractionation of C.

\section{Discussion}

\subsection{Comparison with previous disk models}

The most relevant work on inner disk chemistry is that by
\citet{mar02}, who study the inner 10\,AU of a static disk and assume
that the gas temperature is equal to the dust temperature. Their
physical model differs greatly from ours, although their chemical
network is similar. \citet{mar02} adopt an accretion rate which is a
factor of ten larger than ours -- this will cause their surface
density to be higher, and also their disk to be warmer, in
general. The higher surface density means that fewer photons will
penetrate the disk, and thus the ionisation profile of the disk is
different (their Fig.~3 shows a significantly lower abundance of H$^+$
-- x(H$^+$)$<$10$^{-12}$ -- than our model --
x(H$^+$)$<$10$^{-2}$). Furthermore, the \citet{mar02} model involves a
1\,M$_\odot$ star, 43\% more massive than in our model.  A higher
stellar mass implies a greater stellar gravity and thus a thinner
disk, which receives less stellar UV radiation. Their model also has a
temperature inversion, which means that, surprisingly, they find that
the majority of species are adsorbed onto dust grains in the surface
of the disk at 10\,AU. Given this, it is not surprising that the
predictions of the two models are very different. Ionisation products
such as HCO$^+$ are much more abundant in our model (compare
Fig.~\ref{fig:HCO+dist} with their Fig.~4), and volatile species are
available in the gas phase at different heights and radii due to the
difference in the dust temperature profile caused by the different
accretion rate.

A more recent model from \citet{agu08} treats the chemistry in the
high density PDR-like regions of protoplanetary disks -- the surface
of the disk and the inner 3\,AU. \citet{agu08} use a solely gas-phase
model to investigate the formation of small species such as HCN,
C$_2$H$_2$ and CH$_4$, introducing reactions with significant
activation energy barriers ($\sim$1\,400\,K) which are not included in
chemical networks based on interstellar chemistry, such as the UMIST
and Ohio State networks. The disk model used by \citet{agu08} is
similar to the one used here, and hence comparison is
straightforward. Since they only consider the surface region of the
disk, they calculate column densities vertically from a height
$z_\mathrm{in}$, the height in the disk where the gas is well-shielded
to UV radiation and where all the carbon is as CO. In our disk, this
corresponds to a height of $\sim$3\,$z_h$. Agreement is very good for
small species such as CO, CO$_2$, CH$_4$ and C$_2$H$_2$, although the
model of \citet{agu08} is deficient in H$_2$O, OH, NH$_3$ and HCN
compared to our model. These differences may be explained by different
photodissociation rates and activation barriers for these species. For
instance, the activation barrier for the reaction H$_2$ + CN
$\longrightarrow$ HCN + H is 1\,200\,K in \citet{agu08}, but only
820\,K in this work, with this value taken from the NIST database.


\subsection{Significance for observations}

\begin{figure*}
\includegraphics[width=16cm]{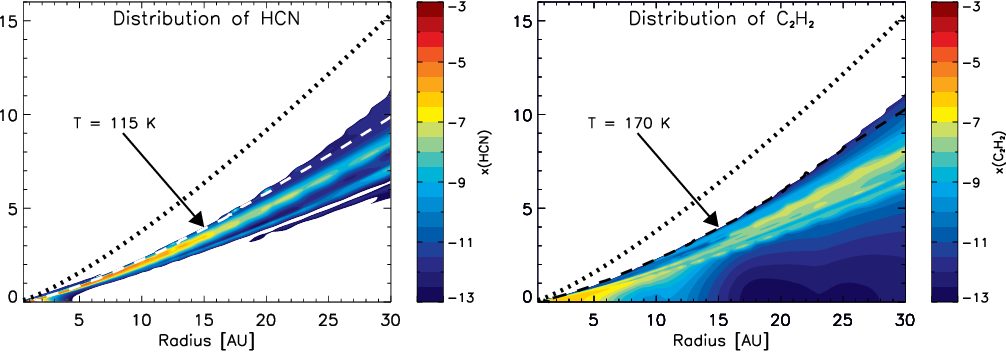}
\caption{The distribution of HCN and C$_2$H$_2$, over-plotted with
isotherms (dashed lines) at the rotational temperatures of these
molecules in GV Tau \citep{gib07} -- 115 and 170\,K,
respectively. }
\label{fig:gib07}
\end{figure*}

Observations of protoplanetary disks at long wavelengths
(sub-millimetre and millimetre) are sensitive to the cold gas of the
midplane and outer regions (R$>$50--100\,AU). Recent observations at
infrared wavelengths are sensitive to warmer gas and dust, and thus
have been able to probe within a few AU of the star. Future
initiatives (e.g., ALMA) which give us high spatial resolution, should
enable us to detect material at the radii of Earth-like planet-forming
regions in other systems.

Some recent observations have led to derivations of the
$^{12}$C/$^{13}$C ratio in one edge-on (\object{GV Tau}) and one
almost face-on (\object{HL Tau}) T Tauri disks, similar to the one we
model. \object{GV Tau} is a binary system in which the T Tauri star
has a stellar mass and effective temperature similar to that which we
use \citep{whi04}. The mass of the disk is estimated to be
0.01\,M$_\odot$ \citep{hog98}. The accretion rate is 20 times greater,
2$\times$10$^{-7}$\,M$_\odot$\,yr$^{-1}$, and the luminosity nearly
2\,L$_\odot$ \citep{whi04}. \object{HL Tau} has a similar accretion
rate and stellar temperature to \object{GV Tau}, but is nearly twice
as massive and slightly less luminous \citep{whi04}. The disk around
\object{HL Tau} extends to at least 200\,AU and has a mass of
$\sim$0.1\,M$_\odot$ \citep{gib04}. Thus there is a good basis for
comparison between our model and these objects.

Through a combination of fundamental and overtone lines of $^{12}$CO
and $^{13}$CO, \citet{gib07} and \citet{bri05} derived
$^{12}$CO/$^{13}$CO ratios of 54$\pm$15 for \object{GV Tau} and
76$\pm$9 for \object{HL Tau}.  These ratios are typical of the
transition region (\object{GV Tau}) and surface regions (\object{HL
Tau}), assuming that these disks have an average $^{12}$C/$^{13}$C
ratio of 77.

\citet{gib07} were able to calculate rotation temperatures for a
number of species in \object{GV Tau}, including HCN (115$\pm$11\,K)
and C$_2$H$_2$ (170$\pm$19\,K), as well as $^{12}$CO (200$\pm$40\,K)
and $^{13}$CO (260$\pm$20\,K). Using these and other constraints,
\citet{gib07} were able to locate the origin of the hydrocarbon
absorption to a region above the midplane, possibly the disk
atmosphere, within $\sim$10\,AU of the central star. These
temperatures tally very well with our model: we have plotted the
distributions of HCN and C$_2$H$_2$ from our model in
Fig.~\ref{fig:gib07}. The contour lines show isotherms at the rotation
temperatures calculated by \citet{gib07}. The agreement is very good,
especially given the very narrow vertical range in which these two
molecules are abundant, due to freezeout in colder layers and either
photodestruction or reaction with reactive ions in the layers
above. \citet{whi04} make an estimate of the luminosity of \object{GV
  Tau} which is larger than that which we use in our model. Taking
this into account might move the isotherm closer to the disk midplane,
thus improving the agreement with the narrow molecule-rich layers.

\citet{gib07} also derived column densities from their observations
through the edge-on disk of \object{GV Tau}. \citet{bri05} calculated
$N$($^{12}$CO)=7.5\,$\times$10$^{18}$\,cm$^{-2}$ and
$N$($^{13}$CO)=9.9\,$\times$10$^{16}$\,cm$^{-2}$ from their
observations of \object{HL Tau}, and these column densities correspond
to a radius of greater than 35\,AU in our model. Derived rotational
temperatures of 105\,K and 80\,K, respectively, indicate that the
probed region is an intermediate layer of the disk.

Methane abundances are also constrained by \citet{gib07,gib04}, but
rotational temperatures are not given. The gas temperature of the
methane-rich layer in our disk is $\approx$100\,K. In terms of column
density, our calculated vertical column densities of
$N$(CH$_4$)=3--21$\times$10$^{17}$\,cm$^{-2}$ at radii in the inner
10\,AU of the disk are a factor of 20--300 above the observed upper
limits for the \object{GV Tau} disk \citep{gib07}. This discrepancy
could be down to differences in orientation between \object{GV Tau}
and our model (column densities of methane and other molecules in our
model are calculated vertically, whereas \object{GV Tau} is edge-on to
the line of sight), or due to uncertainties in the chemistry of
methane \citep[perhaps in the accurancy of activation energy barriers
in the reactions which lead up to methane formation,
see][]{agu08}. Abundances of methane are also dependent on initial
conditions, with methane forming efficiently at the start of the
interstellar cloud model on grains, due to high initial abundances of
atomic C and H. Calculated column densities are given in
Table~\ref{tab:coldens}.

\begin{figure*}
\includegraphics[width=16cm]{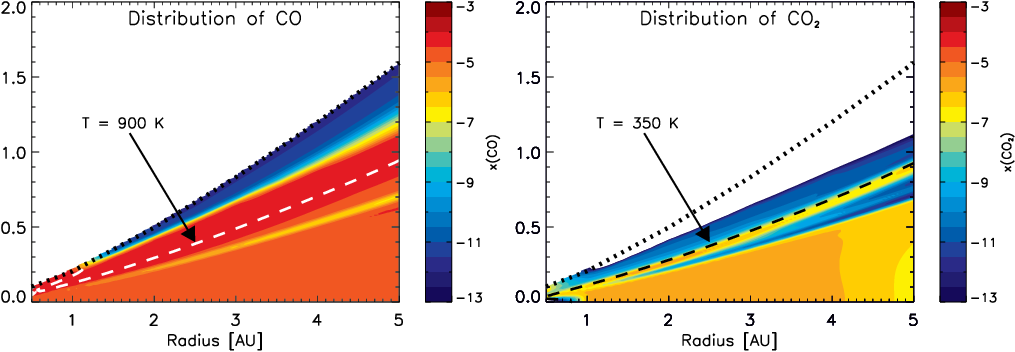}
\includegraphics[width=16cm]{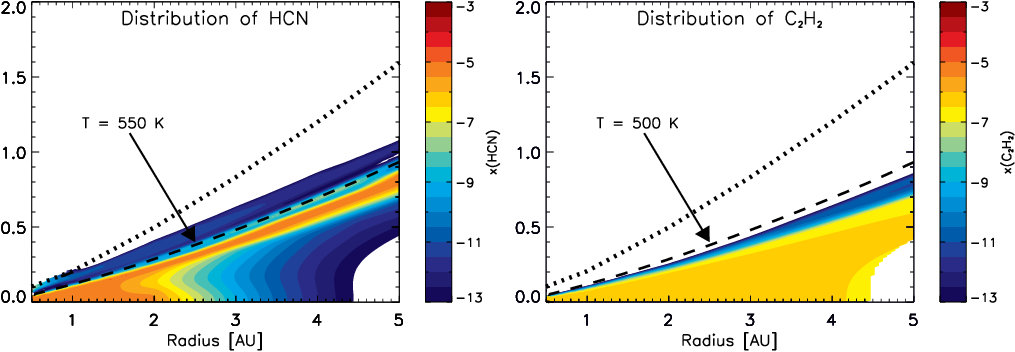}
\includegraphics[width=16cm]{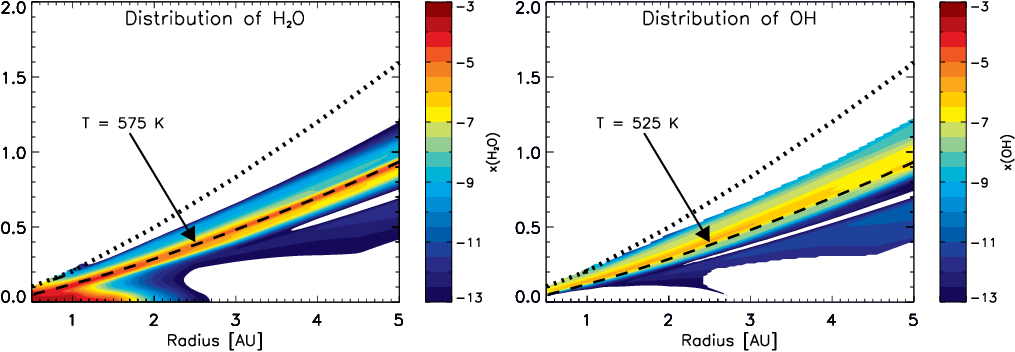}
\caption{The distribution of CO, CO$_2$, HCN, C$_2$H$_2$, H$_2$O and
OH over-plotted with isotherms (dashed lines) at the rotational
temperatures of these molecules in AA Tau \citep{car08} -- 900, 350,
550, 500, 575 and 525\,K, respectively. }
\label{fig:car08}
\end{figure*}

\citet{car08} have probed the inner few AU of the edge-on
circumstellar disk of \object{AA Tau}, a typical classical T Tauri
star, with observations from the Spitzer Space Telescope. They report
detections of the small molecules CO, C$_2$H$_2$, HCN, OH and H$_2$O,
with emission originating from within 2--3\,AU of the star. The
regions from which this molecular emission arises are hot
(525--900\,K). CO$_2$ emission comes from cooler regions (350\,K), and
thus is assumed to come from larger radii. Our model shows very good
correlation between regions of high abundance of these species and gas
temperature with the results of \citet{car08}. There is a slight
disagreement with C$_2$H$_2$, which in our model is found in regions
with temperatures of 70--250\,K, significantly cooler than the 650\,K
calculated by \citet{car08}. See Fig.~\ref{fig:car08}, which shows
distributions of the molecules detected by \citet{car08} and isotherms
at the designated temperatures attributed to those molecules.

\begin{deluxetable}{llcccccccc}
\tablecaption{\label{tab:coldens} Selected vertical column densities of molecules in the
model disk at 1, 10, 20 and 30\,AU}
\tablewidth{0pt}
\tablehead{
\colhead{Species} && \colhead{Column at 1\,AU} && \colhead{Column at 10\,AU} && \colhead{Column at 20\,AU} && \colhead{Column at 30\,AU}}
\startdata
H$_2$             && 8.3\,(24)                 && 1.2\,(24)                  && 5.7\,(23) && 3.7\,(23)\\
He                && 2.3\,(24)                 && 3.3\,(23)                  && 1.6\,(23) && 1.0\,(23)\\
CO                && 3.7\,(20)                 && 5.6\,(19)                  && 3.1\,(19) && 2.4\,(19)\\
CO$_2$            && 2.4\,(19)                 && 3.2\,(14)                  && 8.7\,(12) && 6.3\,(12)\\ 
CH$_4$            && 2.1\,(18)                 && 2.5\,(17)                  && 1.8\,(17) && 4.3\,(18)\\
C$_2$H$_2$        && 1.2\,(19)                 && 3.2\,(15)                  && 7.0\,(13) && 4.9\,(13)\\
C$_3$H$_2$        && 6.0\,(18)                 && 1.4\,(16)                  && 8.7\,(15) && 7.7\,(14)\\
C$_3$H$_3$        && 1.6\,(18)                 && 3.3\,(12)                  && 4.3\,(11) && 2.9\,(12)\\
C$_3$H$_4$        && 6.6\,(19)                 && 2.5\,(16)                  && 9.2\,(11) && 5.0\,(12)\\
C$_4$H$_2$        && 2.1\,(18)                 && 8.1\,(14)                  && 1.1\,(15) && 5.1\,(14)\\
H$_2$O            && 2.3\,(21)                 && 4.3\,(14)                  && 8.1\,(13) && 7.2\,(12)\\
N$_2$             && 1.6\,(19)                 && 2.1\,(17)                  && 1.1\,(17) && 7.5\,(16)\\
NH$_3$            && 2.3\,(20)                 && 1.2\,(15)                  && 3.7\,(13) && 2.0\,(12)\\
HCN               && 8.4\,(19)                 && 1.6\,(15)                  && 2.7\,(13) && 3.7\,(12)\\
HC$_3$N           && 7.0\,(18)                 && 2.9\,(15)                  && 4.2\,(13) && 9.9\,(12)\\
$^{13}$CO         && 7.9\,(18)                 && 1.2\,(18)                  && 6.3\,(17) && 4.4\,(17)\\
$^{13}$CC$_2$H$_4$&& 1.8\,(18)                 && 6.6\,(14)                  && 2.8\,(10) && 1.5\,(11)\\
g--C$_2$H$_2$     && 1.4\,(9)                  && 2.5\,(17)                  && 3.2\,(18) && 1.4\,(18)\\
g--C$_2$H$_6$     && 1.1\,(20)                 && 1.1\,(19)                  && 8.4\,(17) && 2.9\,(17)\\
g--C$_3$H$_2$     && 2.2\,(8)                  && 3.1\,(14)                  && 2.4\,(18) && 2.2\,(18)\\  
g--C$_3$H$_4$     && 8.3\,(9)                  && 1.8\,(19)                  && 6.2\,(18) && 1.4\,(18)\\
g--C$_4$H$_2$     && 1.5\,(10)                 && 1.9\,(17)                  && 1.6\,(18) && 1.1\,(18)\\
g--H$_2$O         && 2.2\,(20)                 && 3.5\,(20)                  && 1.7\,(20) && 1.0\,(20)\\
g--NH$_3$         && 1.9\,(12)                 && 3.8\,(19)                  && 1.8\,(19) && 1.2\,(19)\\
g--HCN            && 1.2\,(15)                 && 9.7\,(18)                  && 4.6\,(18) && 3.0\,(18)\\
g--$^{13}$CCH$_6$ && 1.9\,(18)                 && 2.0\,(17)                  && 1.8\,(16) && 6.1\,(15)\\
\enddata
\tablecomments{$x$\,($y$) here represents
$x\times10^{y}$\,cm$^{-2}$. g-- before a species denotes that that
species is adsorbed onto a grain surface.}
\end{deluxetable}

\subsection{Significance for the Solar System}

\begin{figure*}
\includegraphics[width=16cm]{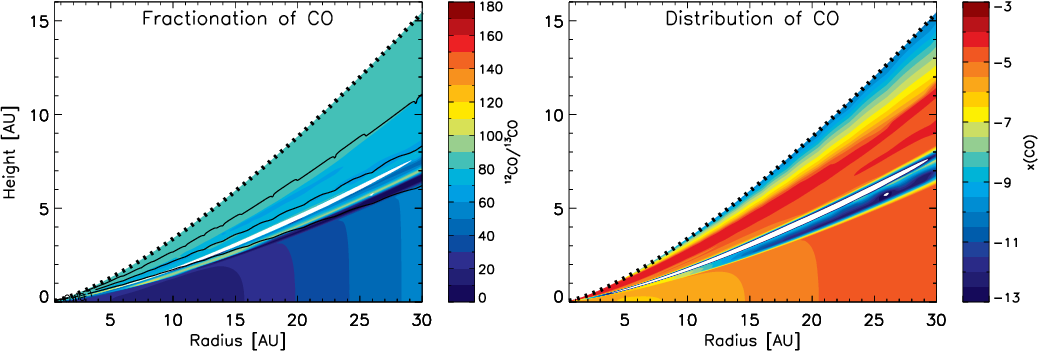}
\caption{Carbon monoxide fractionation and distribution in the MMSN
model. Compare to Fig.~\ref{fig:COfrac}.}
\label{fig:MMSN}
\end{figure*}

Estimates of the mass of the protosolar disk suggest that it was
somewhat more massive than the disk considered in this paper. To make
a better comparison with Solar System data we have therefore also
considered a model with $\dot{M}$=10$^{-8}$\,M$_\odot$\,yr$^{-1}$ and
$\alpha$=0.025, computed for us by Paola D'Alessio. This value of
$\alpha$ results in a disk mass similar to the Minimum Mass Solar
Nebula \citep[MMSN;][]{hay81}. All other parameters are the same as
the fiducial model, apart from a new scaling for $z_h$,
$z_h$=0.0244R$^{18/13}$. We input [$^{12}$C]/[$^{13}$C]=89 to the
molecular cloud model to match present Solar System $^{12}$C/$^{13}$C
isotope ratios.

Results from this model differ from the results of our fiducial model
in two respects. Since the MMSN model is so much more dense, radial
advection times are longer -- a parcel of gas will pass from the outer
edge of the model at 35\,AU into the star in a time of 1.6\,Myr, some
four times greater than the fiducial model. This gives each parcel a
greater processing time at each radius, which leads to:

1) a greater range of degree of fractionation in the disk, such that,
for instance, the fractionation in CO decreases from its input value
into the disk of 50 to a minimum of 11, compared to 44 and 36,
respectively, in the midplane of the fiducial model.

2) a steeper fractionation gradient in the midplane. This increase in
   range of the fractionation occurs over the same distance in the
   model, thus the rate of change of fractionation is greater. This is
   also the case for vertical changes in fractionation, with often
   large changes in fractionation occurring over small distances
   ($\sim$1\,AU).

These differences can be seen in Fig.~\ref{fig:MMSN} for CO.

\begin{figure*}
\includegraphics[width=10cm]{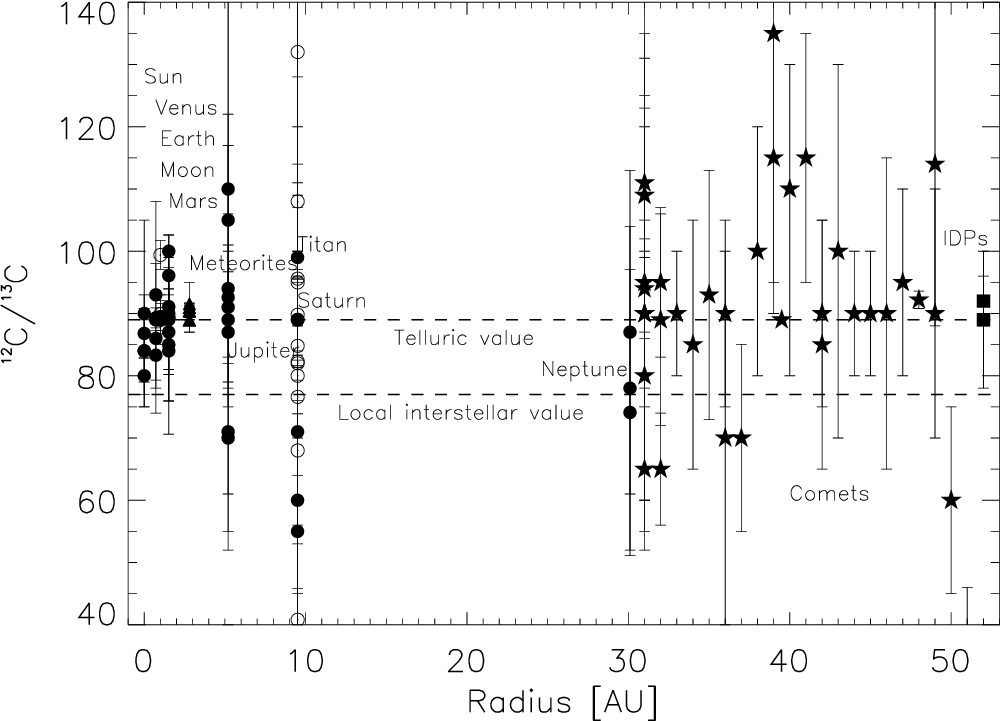}
\caption{Measurements of the $^{12}$C/$^{13}$C ratio in various
objects of the Solar System. Filled circles indicate measurements of
planets or the Sun and empty circles indicate measurements of
planetary moons. Triangles indicate bulk isotope measurements of the
$^{12}$C/$^{13}$C ratio in meteorites, and have been placed at the
radius of the asteroid belt. Comets, indicated by filled stars, have
been placed outside of the radius of Neptune, for illustration, and
similarly, IDPs (filled squares) have been placed at cometary radii to
indicate their likely origin in comets. Please contact the author for
a full list of references for the data in this plot.}
\label{fig:SSiso}
\end{figure*}

Measurements of the $^{12}$C/$^{13}$C ratio in the present day Solar
System cluster around the telluric value of 89
(Fig.~\ref{fig:SSiso}). Although error bars in some cases can be very
large, the majority of measurements are consistent with the idea that
the bulk of Solar System material comes from a common origin, with an
isotope ratio of 89. The most direct comparison we can draw between
our MMSN model and present-day fractionation ratios is in the icy
matter of comets, which is generally considered to be pristine.
Cometary and meteoritic material is thought to be remnant from the
very earliest phases of the formation of the Solar System
\citep[e.g.,][]{mes00}. The degree to which this material has been
processed is unknown, although clues can be found in analyses of
interplanetary dust particles (IDPs) and carbonaceous chondrites (CCs)
on meteorites. IDPs, which become trapped in the Earth's atmosphere,
possibly originate in comets, whilst CCs come from the asteroid
belt. Both types of compound may have been subject to heating, mixing
and chemical reactions during the history of the solar system,
possibly eradicating any chemical ``history''. However, the results of
carbon fractionation measurements show that in general, all measured
comets have a similar ratio (see Fig.~\ref{fig:SSiso}). Similarly,
experiments replicating solar wind or cosmic-ray processing of comet
surfaces has shown that these effects do not significantly contribute
to carbon isotopic fraction \citep{lec98}. This is surprising, since
$^{12}$C/$^{13}$C ratios have been derived from both long- and
short-period comets, which are thought to have formed in different
regions of the protosolar disk.

Carbon fractionation ratios in comets have thus far been determined by
observing (one of) three molecules - HCN, CN and C$_2$ - all members
of the carbon isotope pool. CN is most likely a photodissociation
product of HCN in cometary comae, although other parents may exist
\citep{arp03}. In our MMSN disk model CN and C$_2$ ices are of very
low abundance, which makes it difficult to give a $^{12}$C/$^{13}$C
ratio with a high degree of confidence. However, HCN ice is more
plentiful, and has a $^{12}$C/$^{13}$C ratio of 126--129 in the
midplane region of the disk model. This ratio is inherited from the
molecular cloud. It matches very well with observations of Comet Hale
Bopp, a long-period comet: H$^{12}$CN/H$^{13}$CN = 110$\pm$12
\citep{jew97,ziu99}, but not with Comet Hyakutake:
H$^{12}$CN/H$^{13}$CN = 34$\pm$12 \citep{lis97}. To our knowledge
these are the only three determinations of the H$^{12}$CN/H$^{13}$CN
in comets. In the disk model, CN generally has a similar fractionation
ratio to HCN. However, fractionation ratios derived from observations
of CN in comets are somewhat different to HCN, in the region 65-115,
with an average of $\sim$90. This may be another indication of the
alternative parentage of the CN molecule, or perhaps that there are
some photo-fractionation effects in the photodissociation of HCN.


What conclusions, then, can be drawn from this work given that our
standard model of fractionation in a protoplanetary disk produces a
very ``stratified'' result, where fractionation differs from region to
region in the disk, and also from species to species? One possibility
is that the protosolar disk was heated to some sufficiently high
temperature for chemical exchange reactions to be ineffectual. This
may have had the effect of ``resetting'' the carbon isotope ratios,
similar to the process which occurred later in the formation of
individual planets and the Sun. However, cooling from this state would
have to have been relatively fast, faster than the chemical timescales
for fractionation.

A slightly different possibility relies on mixing material up to the
surface layers of the disk, heating and photodestroying it, leading to
a reset of the the carbon isotope ratio. For the processed material to
be mixed down to the planetary accretion zones at the midplane of the
disk would require vertical mixing timescales to be fast in relation
to both the chemical timescales and accretion timescales. Given the
high densities in the MMSN model, this seems unlikely since collision
times and chemical reaction times will be short.

A third consideration is that of nebular shocks, which may have
transiently heated material in the inner nebula
\citep{des02,kre02}. Processed material could then have rapidly cooled
and frozen out onto grain surfaces within chemical fractionation
timescales, thus locking in the carbon isotope ratio of the hotter
gas. However, the entire nebula would have had to have passed through
such a shock to produce such a uniform carbon isotope ratio.

So we conclude that whatever the mechanism which homogenised carbon
isotope fractionation ratios in the Solar System (and these mechanisms
bear future investigation), it occurred before comets and planets
formed, yet after the initial collapse of the solar nebula's parent
molecular cloud.

\section{Conclusions}

We have shown by means of a chemical model of a protoplanetary disk
that the fractionation ratio of carbon, $^{12}$C/$^{13}$C, varies
according to position in the disk. The fractionation ratio is governed
by temperature, which affects the rate and direction of chemical
exchange reactions, and incident UV radiation, which affects
self-shielding molecules such as CO. This produces a picture of the
fractionation in a disk which is stratified. Certainly in the upper
region of the disk, fractionation timescales are faster than mixing
timescales, meaning that this stratified picture should persist if
mixing were to be considered in the model. We also considered a
Minimum Mass Solar Nebula model in which the chemistry in our
inflowing gas packets has longer to evolve. Extremely good agreement
was seen between observations of Solar System comets and the
fractionation ratio in ices in the midplane of the disk model. In
general, our chemical model has excellent agreement with recent
observations of T Tauri disks, both in terms of chemical abundance and
location in the disk.

\acknowledgements

The authors would like to thank Paola D'Alessio for kindly providing
us with the results of her hydrostatic disk models and Hal Yorke for
allowing us to use his radiative transfer code. We would also like to
thank Neal Turner, Geoff Bryden and Bill Langer at JPL for useful
discussions and the anonymous referee for an extremely thorough
reading of the manuscript. This research was conducted at the Jet
Propulsion Laboratory, California Institute of Technology, under
contract with the National Aeronautics and Space Administration. PMW
was supported by an appointment to the NASA Postdoctoral Program at
JPL, administered by ORAU through a contract with NASA. Partial
support was provided to KW by the NASA Terrestrial Planet Finder
Foundation Science Program.

\appendix

\section{Gas heating and cooling mechanisms}
\label{sec:heatcool}

\subsection{Heating}
\label{sec:heating}

\paragraph{Photoelectric heating.}
Photoelectrons are emitted from dust grains when UV photons impinge
upon the grain. The energy of these emitted particles depends on the
energy of the incident photon and on the emitting grain potential. The
gas heating rate due to photoelectrons is approximated by
\citet{bak94}, and is implemented as follows \citep{kam01}.

For the purposes of the thermal balance, we assume that there are two
populations of dust grain within the disk -- one consisting of
silicate particles and the other consisting of a mixture of small
graphitic and PAH grains. The silicate grains are assumed to have a
radius, $a=0.1$\,$\mu$m, and a UV cross section per H nucleus
($\sigma_\mathrm{UV}$) of $5.856\times10^{-22}$\,cm$^2$/H atom
\citep{kam04}. The small graphite grains are assumed to have a size
distribution ranging from 3--100\AA \citep{bak94}. Larger grains do
not contribute to the disk heating.

Thus the photoelectric heating rate, $\Gamma_\mathrm{PE}$, is given
by:
\begin{equation}
\Gamma_\mathrm{PE} = b\epsilon\chi n_0\units
\end{equation}
\citep{bak94} where $\epsilon$ is an efficiency factor depending on
the material of the grain, $\chi$ is the photon flux measured in units
of the \citet{hab68} field between 912-1110\,\AA\ attenuated by dust
and $n_0$ is the total number density of hydrogen nuclei,
$n_0$=2$n_\mathrm{H_2}$+$n_\mathrm{H}$. $b$ is a constant and is equal
to $1\times10^{-24}$ for graphite/PAH grains and
$2.5\times10^{-4}\sigma_\mathrm{UV}$ for silicate grains, with
$\sigma_\mathrm{UV} = Q_\mathrm{abs}\pi a^2 n_\mathrm{d}$, where
$Q_\mathrm{abs}$ is a UV absorption efficiency and $n_\mathrm{d}$ is
the number density of dust grains.

The photoelectric efficiencies and yields depend on
material. Following \citet{kam01}, we define a grain charge parameter,
$x\equiv\chi T^{0.5}/n_\mathrm{e}$, with $n_\mathrm{e}$ the electron
density. Thus,
\begin{eqnarray}
\epsilon_\mathrm{sil} &=& \frac{6\times10^{-2}}{1 + 1.8\times10^{-3}x^{0.91}} + 
\frac{1.6\times10^{-5}y_\mathrm{sil}T^{1.2}}{1 + 1\times10^{-2}x}\\
\epsilon_\mathrm{PAH} &=& \frac{0.0487}{1 + 4.0\times10^{-3}x^{0.73}} + 
\frac{5.8\times10^{-5}T^{0.7}}{1 + 2\times10^{-4}x}.
\end{eqnarray}
\citep{kam01}. The yields for silicate material ($y_\mathrm{sil}$) are:
\begin{eqnarray}
y_\mathrm{sil} = \left\{ \begin{array}{cl}
                        0.70 & \mathrm{for~} x\leq 10^{-4}\\
                        0.36 & \mathrm{for~} 10^{-4} < x \leq 1\\
                        0.15 & \mathrm{for~} x > 1
			\end{array}\right.
\end{eqnarray}

\paragraph{Collisional de-excitation of H$_2$.}
\label{sec:colldex}
H$_2$ molecules in excited rovibrational levels (due to Lyman-Werner
band absorptions, for example) can decay to less energetic levels
following collisions, thereby heating the gas in the process. This
complex situation was simplified by \citet{tie85} by considering a
single excited pseudovibrational level of H$_2$. They derived the
following heating rate:
\begin{equation}
\Gamma_\mathrm{CDx} = (n_\mathrm{H}\gamma_\mathrm{*0}^\mathrm{H} +
n_\mathrm{H_2}\gamma_\mathrm{*0}^\mathrm{H_2})n_\mathrm{H_2^*}E_*\units,
\end{equation}
where $n_\mathrm{H}$ is the number density of atomic hydrogen,
$n_\mathrm{H_2}$ is the number density of molecular hydrogen,
$n_\mathrm{H_2^*}$ is the number density of vibrationally excited
H$_2$ (presumed to be a fixed proportion of H$_2$,
$n_\mathrm{H_2^*}$=$10^{-5}n_\mathrm{H_2}$) and $E_*$ is the effective
energy of the pseudolevel \citep[taken to be
$4.166\times10^{-12}$\,ergs,][]{lon78}. $\gamma_\mathrm{*0}^\mathrm{H}$
and $\gamma_\mathrm{*0}^\mathrm{H_2}$ are the collisional
de-excitation rate coefficients from the excited $v$=6 level to the
$v$=0 level. Since these transitions tend to occur in steps of $\Delta
v$=1, $\gamma_\mathrm{*0}^\mathrm{H}$ and
$\gamma_\mathrm{*0}^\mathrm{H_2}$ are assumed to equal to one sixth of
the $v$=1-0 rate coefficients \citep{tie85}. Thus
$\gamma_\mathrm{*0}^\mathrm{H} =
1.67\times10^{-13}\sqrt{T}\exp(-1\,000\,\mathrm{K}/T)$\,cm$^3$\,s$^{-1}$ and
$\gamma_\mathrm{*0}^\mathrm{H_2} =
2.33\times10^{-13}\sqrt{T}\exp(-18\,100\,\mathrm{K}/[T+1\,200\,\mathrm{K}])$\,cm$^3$\,s$^{-1}$.

\paragraph{Photodissociation of H$_2$.}
Whilst 90\% of Lyman-Werner band excitation results in collisional
de-excitation, the remaining 10\% results in radiative dissociation to
a pair a hydrogen atoms, each carrying approximately 0.4\,eV of
kinetic energy \citep{ste73}. Hence the heating rate due to
photodissociation of H$_2$ can be approximated:
\begin{equation}
\Gamma_\mathrm{Phd} = 5.55\times10^{-13}\chi\Gamma_\mathrm{H_2}^\prime
n_\mathrm{H_2}\units
\end{equation}
\citep{tie85}. Here $\Gamma_\mathrm{H_2}^\prime$ is the
photodissociation rate of H$_2$, taking self-shielding into account
(see Sect.~\ref{sec:selfshield}).

\paragraph{H$_2$ formation.}
The formation and release of a hydrogen molecule from the surface of a
dust grain contributes 4.48\,eV of binding energy to the thermal
balance. Since it is unclear how much of that energy is transferred
into rotation, vibration and translation, we follow the approach of
\citet{bla76} \citep[and also][]{kam01} by assuming that one third
goes into each motion. This results in a heating rate due to H$_2$
formation:
\begin{equation}
\Gamma_\mathrm{Form} = 2.39\times10^{-12}R_\mathrm{form}\units.
\end{equation}
$R_\mathrm{form}$ is the formation rate of H$_2$ taken from
\citet{caz02a}, which incorporates revised H$_2$ formation efficiencies
($\epsilon_\mathrm{H_2}$) at high temperatures,
viz.
\begin{eqnarray}
\label{h2form}R_\mathrm{form} &=&
0.5n(\mathrm{H})v_\mathrm{H}n_\mathrm{d}\sigma_\mathrm{d}\epsilon_\mathrm{H_2}S_\mathrm{H}\quad\mathrm{cm^{-3}\,s^{-1}}\\
\epsilon_\mathrm{H_2} &=& \left(1 + \frac{\mu F}{2\beta_\mathrm{H_2}}
+ \frac{\beta_\mathrm{H_p}}{\alpha_\mathrm{pc}}\right)^{-1}\xi\\ \xi
&=& \left[1 +
  \frac{\nu_\mathrm{H_c}}{2F}\exp\left(-\frac{1.5E_\mathrm{H_c}}{kT}\right)\left(1
  +
  \sqrt{\frac{E_\mathrm{H_c}-E_\mathrm{s}}{E_\mathrm{H_p}-E_\mathrm{s}}}\right)^2\right]^{-1}.\label{Eq:xi}
\end{eqnarray}
$v_\mathrm{H}$ is the velocity of a hydrogen atom and
$\sigma_\mathrm{d}$ is the geometric cross section of a
grain. $n_\mathrm{d}$ is usually assumed to be $\approx$10$^{-12}n_0$,
but can be calculated from the gas density:
\begin{equation}
\label{eq:nd} n_\mathrm{d} = \delta_\mathrm{dg}n_0\mu
m_\mathrm{H}\left(\frac{4\pi a^3
\rho_\mathrm{d}}{3}\right)^{-1}\quad\mathrm{cm}^{-3}.
\end{equation}
We adopt a standard dust-to-gas mass ratio ($\delta_\mathrm{dg}$) of
0.01, a grain density ($\rho_\mathrm{d}$) of 2.5\,g\,cm$^{-3}$ and a
grain size ($a$) of 0.1$\mu$m. $\mu$ is the reduced mass of the gas
(taken to be 2.4) and $m_\mathrm{H}$ the mass of an H
atom. $S_\mathrm{H}$ in Eq.~\ref{h2form} is the sticking coefficient
of hydrogen and is assumed to be 0.4, in line with \citet{caz04}.  The
reader is referred to \citet{caz02b,caz02a} for an explanation of the
other terms. In the calculation of $\epsilon_\mathrm{H_2}$ we assume
$F$=10$^{-15}$ (Kamp, priv. comm.), where $F$ is the accretion rate of
H$_2$ in units of monolayers per second. In the calculation of $\xi$,
we assume that when a grain is completely covered with a monolayer of
ice no chemisorption of H atoms can occur, but physisorption can. Thus
in Eq.~\ref{Eq:xi}, $E_\mathrm{H_c}$=$E_\mathrm{H_p}$=600\,K, rather
than $E_\mathrm{H_c}$=10\,000\,K in the chemisorption case.

\paragraph{C ionisation.}
Neutral carbon can be readily photoionised in protoplanetary disks,
with the release of approximately 1\,eV of energy per ionisation
\citep{bla87}. The heating rate due to this process depends on the
attenuation by dust absorption \citep{bla77}, attenuation by
self-absorption of C \citep{wer70} and also attenuation by the H$_2$
column \citep{dej80} (the first, second and third exponential terms,
respectively).
\begin{equation}
\Gamma_\mathrm{CIon} =
2.2\times10^{-22}n(\mathrm{C})\chi_0\frac{\exp(-2.4A_\mathrm{V} -
  \tau_\mathrm{C} - \tau_\mathrm{H_2} v_2/\pi v_1^2)}{1 +
  \tau_\mathrm{H_2} v_2/\pi v_1^2}\units
\end{equation}
\citep{tie85}. The various parameters are given by:
\begin{eqnarray}
\nonumber\tau_\mathrm{C}=10^{-17}N(\mathrm{C}),&\quad&\tau_\mathrm{H_2}=1.2\times10^{-14}N(\mathrm{H_2})\dvd,\\
\quad v_2=9.2\times10^{-3}\dvd,&\quad&v_1=5\times10^2\dvd.
\end{eqnarray}
$\chi_0$ is the unattenuated photon flux incident on the surface of
the disk, $N(\mathrm{H_2})$ and $N(\mathrm{C})$ are the column
densities of molecular hydrogen and atomic carbon,
respectively. $\delta v$ is the line broadening, taken to be equal to
the sound speed.

\paragraph{Cosmic ray heating.}
Cosmic rays will penetrate the disk up until a certain column density
of matter \citep[150\,g\,cm$^{-2}$;][]{ume81}, and contribute to the
thermal balance via the energy released in the formation and
subsequent electronic recombination of H$_3^+$. This process yields
some 7\,eV per ionisation \citep{gla73}. The heating rate is given by:
\begin{equation}
\Gamma_\mathrm{C-ray} = (1 + n_\mathrm{He}/n_0)
\zeta_\mathrm{H}n_0(1.28\times10^{-11} +
2.44\times10^{-11}(n_\mathrm{H_2}/n_0))\units
\end{equation}
\citep{cla78}, where $n_\mathrm{He}$ is the number density of
helium. For $\zeta_\mathrm{H}$, the cosmic ray ionisation rate of
hydrogen, we use the value of 5.98$\times$10$^{-18}$\,s$^{-1}$
\citep{umist06}.

\paragraph{X-ray heating.}
To calculate the heating effect due to X-rays we use the prescription
of \citet{gor04} for an X-ray energy spectrum of 0.5-10\,keV:
\begin{equation}
\label{eq:Xrayheat}
\Gamma_\mathrm{X-ray} =
\int_{0.5}^{10}F(E)\exp(-N_0\sigma_\mathrm{X}(E))\sigma_\mathrm{X}(E)n_0f_\mathrm{heat}dE\units.
\end{equation}
$F(E)$ is the X-ray photon flux at a radius R: $L(E)$/$4\pi R^2$,
where $L(E)$ is the stellar luminosity as a function of emitted energy
(in keV). \citet{gor04} fit a broken power law to the X-ray spectrum
of a weak-line T Tauri star presented in \citet{fei99}, resulting in:
\begin{eqnarray}
\label{eq:Xrayspec}
L(E) = \left\{ \begin{array}{cl}
       1.2L_\mathrm{X}E^{-1.75}\quad & \mathrm{for~} E>2\,\mathrm{keV},\\
       0.18L_\mathrm{X}E & \mathrm{for~} E<2\,\mathrm{keV}.
       \end{array}\right.
\end{eqnarray}
$L_\mathrm{X}$ is the X-ray luminosity of the central star in
ergs\,s$^{-1}$, which is in general 10$^{-3}$--$10^{-4}\,L_\star$
\citep{fei99}. We use $L_\mathrm{X}$=$10^{-4}\,L_\star$, and use the
fit in Eq.~\ref{eq:Xrayspec}, since the X-ray spectrum of a classical
T Tauri star resembles that of a weak-line T Tauri star quite closely
\citep{fei99}. In Eq.~\ref{eq:Xrayheat}, $N_0$ is the hydrogen nucleus
column density of gas towards the central star, and
$\sigma_\mathrm{X}(E)$ is the total X-ray photoabsorption
cross-section per H nucleus:
\begin{equation}
\sigma_\mathrm{X}(E) = 2.27\times10^{-22}(E/1\,\mathrm{keV})^{-2.485}\quad\mathrm{cm^2/H~nucleus},
\end{equation}
\citep{wilms00}. $f_\mathrm{heat}$ is the fraction of absorbed energy
which heats the gas, equal to 0.1 for atomic gas and 0.4 for molecular
gas \citep{mal96}.

There is also a secondary X-ray heating effect since the recombination
of H$^+$ after ionisation results in a substantial population of H
atoms in an excited state, which decay collisionally. Thus the primary
rate (Eq.~\ref{eq:Xrayheat}) is augmented by an additional:
\begin{equation}
\Gamma_\mathrm{X-ray}^\prime = 2.22\zeta_\mathrm{X}^\mathrm{H} n_\mathrm{H}
E_{21}\units
\end{equation}
\citep{sha02}. $\zeta_\mathrm{X}^\mathrm{H}$ is the X-ray ionisation
rate of atomic hydrogen and $E_{21}$ the energy gap between the $n$=2
and $n$=1 excited levels of a hydrogen atom
(10.19\,eV). $\zeta_\mathrm{X}^\mathrm{H}$ is calculated according to
the process described in Sect.~\ref{sec:chemistry}.

\subsection{Cooling}
\label{sec:cooling}

The gas in a PPD is cooled by the line emission of atomic and
molecular species, and also, when the gas temperature is greater than
the dust temperature, by collisions between gas and dust.

For the emission lines, we generally make use of the escape
probability factor, $\beta(\tau)$. This factor takes into account the
fact that atomic and molecular lines can have a much higher optical
depth than that of the dust continuum. The escape probability factor,
then, is the probability that a photon from a particular line with an
optical depth, $\tau$, will escape from the disk. The maximum
probability is 0.5, since we assume that any photons emitted in the
negative $z$ direction will be absorbed in the disk.

\citet{tie85} show that the optical depth in the vertical direction
$z$ is:
\begin{equation}
\label{eq:escape}
\tau(z) = A_{ul}\frac{c^3}{8\pi\nu^3(\delta
v)}\int_0^zn_u(z')\left(\frac{n_l(z')g_u}{n_u(z')g_l}-1\right)dz',
\end{equation}
where $z$=0 is the surface of the disk, contrary to our usual
notation. In Eq.~\ref{eq:escape}, $A_{ul}$ is the Einstein transition
probability from level $u$ to level $l$, $\nu$ is the line frequency,
$\delta v$ is the line broadening (assumed to be equal to the sound
speed, $c_\mathrm{S}$), $n_u$ and $n_l$ are the level populations of
levels $u$ and $l$, and $g_u$ and $g_l$ are the corresponding
statistical weights. Ideally, the optical depth would be calculated by
solving the level population equations in non-local thermal
equilibrium (non-LTE). However, this is computationally expensive, and
we have had to make assumptions in order to simplify the
calculation. Hence we assume a plane-parallel slab, with uniform
temperature and density. Level populations are in LTE. The optical
depth then becomes:
\begin{equation}
\label{eq:simpletau}
\tau(z) = \frac{A_{ul}c^3n_1z}{8\pi\nu^3(\delta
v)}\left[\exp\left(\frac{h\nu_{ul}}{kT}\right)-1\right],
\end{equation}
\citep{tie85}. 

The escape probability formalism is defined by \citet{dej80}:
\begin{eqnarray}
\beta(\tau) = \left\{ \begin{array}{cl}
       \frac{1-\exp(-2.34\tau)}{4.68\tau}\quad & \mathrm{for~}
       \tau<7,\\
       \left[4\tau\left(\ln\frac{\tau}{\sqrt{\pi}}\right)^{0.5}\right]^{-1}\quad
       & \mathrm{for~} \tau\geq7.
       \end{array}\right.
\end{eqnarray}

\paragraph{Atomic line cooling.}
To calculate electronic level populations, we solve the statistical
equilibrium equations for a three-level system:
\begin{equation}
n_i\sum_{j\neq i}R_{ij} = \sum_{j\neq
i}n_jR_{ji}\qquad\mathrm{and}\qquad n_\mathrm{X} =
\sum_{j=0}^2 n_j,
\end{equation}
where $n_\mathrm{X}$ is the number density of species X and 
\begin{eqnarray}
R_{ij} = \left\{ \begin{array}{cl}
       A_{ij}\beta(\tau_{ij})(1+Q_{ij}) + C_{ij} \quad & \mathrm{for~}
       i>j,\\
       (g_j/g_i)A_{ji}\beta(\tau_{ji})Q_{ji} + C_{ji},
       & \mathrm{for~} i<j,
       \end{array}\right.
\end{eqnarray}
with
\begin{eqnarray}
\label{eq:finest}
Q_{ij} = Q_{ji} &=& \frac{c^2}{2h\nu_{ij}^3}P(\nu_{ij})\\
P(\nu_{ij}) = B(\nu_{ij},2.7\,\mathrm{K}) &+& B(\nu_{ij},T_\mathrm{d})\tau_d(\nu_{ij}).
\end{eqnarray}
where $C_{ij}$ is the collisional rate from level $i$ to level $j$ and
$P(\nu_{ij})$ denotes the background radiation due to 1) the 2.7\,K
microwave background, and 2) the infrared emission of dust, expressed
in terms of Planck blackbody functions, with $T_\mathrm{d}$ the dust
temperature and $\tau_d(\nu_{ij})$=0.001 \citep{hol91}. Assuming that
the local radiation field in the disk can be represented by the sum of
these two blackbodies is a simplification: it ignores the significant
effect of stellar radiation in the disk surface and neglects the fact
that in the disk midplane the local radiation is optically thick, and
thus not proportional to $\tau_d(\nu_{ij})$. This assumption will
affect the cooling rate and thus the gas temperature \citep[see, for
example][]{kam01}.

We consider the fine-structure lines of [O\textsc{i}], [C\textsc{i}]
and [C\textsc{ii}], in the manner described above in
Eqs.~\ref{eq:simpletau}--\ref{eq:finest}, calculating the statistical
equilibrium in detail, since some regions of the disk are below the
critical density required for LTE to be a reasonable assumption.

The lowest three fine structure lines for [O\textsc{i}] are at 63.2,
145.6 and 44.0\,$\mu$m. Line data is taken from Table 1 of
\citet{kam01}, who take into account collisions with H$_2$, H and
electrons. The cooling rate over all three lines is:
\begin{equation}
\Lambda_{\mathrm{O}\textsc{i}} =
\sum_k\beta(\tau_{ul})h\nu_{ul}[n_u(\mathrm{O})(A_{ul}+B_{ul}P(\nu_{ul})) -
n_l(\mathrm{O})B_{lu}P(\nu_{ul})]\units.
\end{equation}
\citep{kam01}. $h$ is the Planck constant, $\nu_{ul}$ the frequency of
the fine-structure line, $n_u$(O) and $n_l$(O) are the number
densities of oxygen in the upper and lower levels of the transition,
respectively, and $A_{ul}$, $B_{ul}$ and $B_{lu}$ are the Einstein
coefficients for the transition.

A similar approach is taken for [C\textsc{i}], although the critical
density for the three [C\textsc{i}] cooling lines at 609.2, 229.9 and
369.0\,$\mu$m is much lower, and LTE can be assumed. Again, atomic
data is taken from \citet{kam01}. The cooling rate is simply:
\begin{equation}
\Lambda_{\mathrm{C}\textsc{i}} =
h[A_{10}\beta(\tau_{10})\nu_{10}n_1(\mathrm{C}) +
A_{20}\beta(\tau_{20})\nu_{20}n_2(\mathrm{C}) +
A_{21}\beta(\tau_{21})\nu_{21}n_2(\mathrm{C})]\units,
\end{equation}
\citep{kam01}.

We also consider the [C\textsc{ii}] line at 157.7\,$\mu$m, again
assuming LTE since the density in the disk is always above the
critical density for this transition. The cooling rate is:
\begin{equation}
\Lambda_{\mathrm{C}\textsc{ii}} = A_{10}\beta(\tau_{10})h\nu_{10}n_1(\mathrm{C^+})\units,
\end{equation}
with atomic data from \citet{kam01}.

\paragraph{Molecular line cooling.}
H$_2$ rotational/vibrational and CO rotational line cooling are the
main molecular cooling lines in the disk, although we have also taken
into account cooling from CH molecules.

The H$_2$ cooling function for the lowest 51 rovibrational energy
levels has been derived by \citet{leb99}, taking into account
collisions with H, He and H$_2$, but not including pumping by UV and
X-ray radiation which will affect the H$_2$ population in high energy
levels. It is only strictly valid between temperatures of
10$^2$--10$^4$\,K and densities of 1--10$^8$\,cm$^{-3}$. H$_2$ cooling
is negligible at densities less than 10$^6$\,cm$^{-3}$ \citep{hol79}
and at temperatures less than 100\,K, and we persist in using the
function at the high densities which can be found in the disk. We
adopt an ortho-to-para ratio of 1.

26 rotational lines of CO are considered in calculating the cooling
rate due to CO emission:
\begin{equation}
\Lambda_\mathrm{CO} = \beta(\tau_\mathrm{CO})
\sum_{i=1}^{25}h\nu_{ij}[n_i(\mathrm{CO})(A_{ij}+B_{ij}P(\nu_{ij}))-n_j(\mathrm{CO})B_{ji}P(\nu_{ij})]\units.
\end{equation}
Collisional rate coefficients are from \citet{sch85}, and other
molecular data from \citet{kam01}. The optical depth of CO lines is
calculated using Eq.~\ref{eq:escape}.

The emission lines of CH radicals have a minimal effect on the thermal
balance in the inner disk, but can provide cooling in regions where CH
is particularly abundant. We use the cooling rate of \citet{kam01},
\begin{equation}
\Lambda_\mathrm{CH} = \beta(\tau_\mathrm{CH})n_0n_\mathrm{CH}L_\mathrm{rot}\units,
\end{equation}
where the line cooling coefficient, $L_\mathrm{rot}$, is derived by
\citet{hol79}:
\begin{eqnarray}
L_\mathrm{rot} = \left\{ \begin{array}{cl}
       \frac{4(kT)^2A_0}{n_0E_0(1+(n_\mathrm{cr}/n_0)+1.5\sqrt{n_\mathrm{cr}/n_0})}\quad&
       \mathrm{erg\,cm^3\,s^{-1}~for~} n_0\gg n_\mathrm{cr},\\
       \frac{kT(1-(n_\mathrm{H_2}/n_0))\sigma_\mathrm{tot}v_\mathrm{T}}{(1+(n_0/n_\mathrm{cr})+1.5\sqrt{n_0/n_\mathrm{cr}})}\quad&
       \mathrm{erg\,cm^3\,s^{-1}~for~} n_0\ll n_\mathrm{cr},
       \end{array}\right.
\end{eqnarray}
and $\beta(\tau_\mathrm{CH})$=1, which overestimates the cooling rate.
The critical density for CH, $n_\mathrm{cr}$, is
6.6$\times$10$^9\sqrt{(T/1\,000\,\mathrm{K})}$\,cm$^{-3}$, and $A_0$
and $E_0$ are 7.7$\times$10$^{-3}$\,s$^{-1}$ and
2.76$\times$10$^{-15}$\,erg respectively. The total inelastic
cross-section of CH, $\sigma_\mathrm{tot}$, is taken to be
1$\times$10$^{-15}$\,cm$^2$ \citep{hol79}. $v_\mathrm{T}$ is the
thermal velocity of the colliding hydrogen atoms.

\paragraph{Cooling at high temperatures.}
At high temperatures (more than several hundred degrees Kelvin)
Ly$\alpha$ and the line emission from the metastable $^1$D--$^3$P
transition of atomic oxygen at 6300\,\AA\ efficiently cool the
gas. Cooling rates for both these processes are given by
\citet{ste89}:
\begin{eqnarray}
\Lambda_\mathrm{Ly\alpha} &=&
7.3\times10^{-19}\beta(\tau_\mathrm{Ly\alpha})n_\mathrm{e}n_\mathrm{H}\exp(-118\,400\,\mathrm{K}/T)\units,\\
\nonumber\Lambda_\mathrm{O6300} &=&
1.8\times10^{-24}\beta(\tau_\mathrm{O6300})n_\mathrm{O}\exp(-22\,800\,\mathrm{K}/T) \times \\
&~&\qquad\qquad\left[n_\mathrm{H}+n_\mathrm{H_2}+\frac{410n_\mathrm{e}}{(9T^{1/2}+6.3\times10^{-8}n_\mathrm{e}T)}\right]\units,\label{eq:O6300}
\end{eqnarray}
with $n_\mathrm{O}$ the number density of oxygen. The third term in
the brackets in Eq.~\ref{eq:O6300} takes into account excitation of
the oxygen atoms by impacting electrons, as described in
\citet{dra83}. $\beta(\tau_\mathrm{Ly\alpha})$=$\beta(\tau_\mathrm{O6300})$=1.

\paragraph{Gas-grain collisions.}
Collisions between dust grains and atoms and molecules will generally
act as an energy transfer mechanism, and will cool the gas if the gas
temperature is larger than the dust temperature, which is generally
the case in the regions under consideration. The heating rate due to
these collisions is:
\begin{equation}
\Lambda_\mathrm{G-G} = 4.0\times10^{-12}(\pi a^2)n_0n_\mathrm{d}\alpha_\mathrm{T}\sqrt{T}(T-T_\mathrm{d})\units,
\end{equation}
where $a$=0.1\,$\mu$m, the thermal accommodation coefficient,
$\alpha_\mathrm{T}$=0.3 \citep{bur83} and $n_\mathrm{d}$ is defined
previously (Eq.~\ref{eq:nd}).

\clearpage



\clearpage





\begin{thebibliography}{123}
\expandafter\ifx\csname natexlab\endcsname\relax\def\natexlab#1{#1}\fi

\bibitem[Ag{\'u}ndez et al.(2008)]{agu08} Ag{\'u}ndez, M., Cernicharo,
J., \& Goicoechea, J.~R.\ 2008, \aap, 483, 831

\bibitem[Aikawa \& Nomura(2006)]{aik06} Aikawa, Y., \& Nomura, H.\
2006, \apj, 642, 1152

\bibitem[Aikawa \& Herbst(2001)]{aik01} Aikawa, Y., \& Herbst, E.\
2001, \aap, 371, 1107

\bibitem[{{Aikawa} \& {Herbst}(1999)}]{a+h99a}
{Aikawa}, Y. \& {Herbst}, E. 1999, \aap, 351, 233

\bibitem[Aikawa et al.(1999)]{aik99} Aikawa, Y., Umebayashi, 
T., Nakano, T., \& Miyama, S.~M.\ 1999, \apj, 519, 705 

\bibitem[{{Anders} \& {Grevesse}(1989)}]{and89}
{Anders}, E. \& {Grevesse}, N. 1989, \gca, 53, 197

\bibitem[{{Andrews} \& {Williams}(2007)}]{and07}
{Andrews}, S.~M. \& {Williams}, J.~P. 2007, \apj, 659, 705

\bibitem[Arpigny et al.(2003)]{arp03} Arpigny, C., Jehin, E.,
Manfroid, J., Hutsem{\'e}kers, D., Schulz, R., St{\"u}we, J.~A.,
Zucconi, J.-M., \& Ilyin, I.\ 2003, Science, 301, 1522

\bibitem[{{Artymowicz} \& {Clampin}(1997)}]{art97}
{Artymowicz}, P. \& {Clampin}, M. 1997, \apj, 490, 863

\bibitem[{{Ayres} {et~al.}(2006){Ayres}, {Plymate}, \& {Keller}}]{ayr06}
{Ayres}, T.~R., {Plymate}, C., \& {Keller}, C.~U. 2006, \apjs, 165, 618

\bibitem[{{Bakes} \& {Tielens}(1994)}]{bak94}
{Bakes}, E.~L.~O. \& {Tielens}, A.~G.~G.~M. 1994, \apj, 427, 822


\bibitem[Bensch et al.(2001)]{ben01} Bensch, F., Pak, I., 
Wouterloot, J.~G.~A., Klapper, G., \& Winnewisser, G.\ 2001, \apjl, 562, L185 

\bibitem[{{Bergin} {et~al.}(2003){Bergin}, {Calvet}, {D'Alessio}, \&
  {Herczeg}}]{ber03}
{Bergin}, E., {Calvet}, N., {D'Alessio}, P., \& {Herczeg}, G.~J. 2003, \apjl,
  591, L159


\bibitem[{{Bisschop} {et~al.}(2006){Bisschop}, {Fraser}, {{\"O}berg}, {van
  Dishoeck}, \& {Schlemmer}}]{bis06}
{Bisschop}, S.~E., {Fraser}, H.~J., {{\"O}berg}, K.~I., {van Dishoeck}, E.~F.,
  \& {Schlemmer}, S. 2006, \aap, 449, 1297


\bibitem[{{Black}(1987)}]{bla87}
{Black}, J.~H. 1987, in Astrophysics and Space Science Library, Vol. 134,
  Interstellar Processes, ed. D.~J. {Hollenbach} \& H.~A. {Thronson}, Jr.,
  731--744

\bibitem[{{Black} \& {Dalgarno}(1976)}]{bla76}
{Black}, J.~H. \& {Dalgarno}, A. 1976, \apj, 203, 132

\bibitem[{{Black} \& {Dalgarno}(1977)}]{bla77}
---. 1977, \apjs, 34, 405

\bibitem[{{Blake} \& {Boogert}(2004)}]{bla04}
{Blake}, G.~A. \& {Boogert}, A.~C.~A. 2004, \apjl, 606, L73

\bibitem[Boogert et al.(2000a)]{boo00a} Boogert, A.~C.~A.,
  Tielens, A.~G.~G.~M., Ceccarelli, C., Boonman, A.~M.~S., van
  Dishoeck, E.~F., Keane, J.~V., Whittet, D.~C.~B., \& de Graauw,
  T.\ 2000, \aap, 360, 683

\bibitem[Boogert et al.(2000b)]{boo00b} Boogert, A.~C.~A., et
  al.\ 2000, \aap, 353, 349

\bibitem[Boogert et al.(2002)]{boo02} Boogert, A.~C.~A., 
Blake, G.~A., \& Tielens, A.~G.~G.~M.\ 2002, \apj, 577, 271 

\bibitem[{{Brittain} {et~al.}(2005){Brittain}, {Rettig}, {Simon}, \&
  {Kulesa}}]{bri05}
{Brittain}, S.~D., {Rettig}, T.~W., {Simon}, T., \& {Kulesa}, C. 2005, \apj,
  626, 283

\bibitem[{{Brittain} {et~al.}(2003){Brittain}, {Rettig}, {Simon}, {Kulesa},
  {DiSanti}, \& {Dello Russo}}]{bri03}
{Brittain}, S.~D., {Rettig}, T.~W., {Simon}, T., {Kulesa}, C., {DiSanti},
  M.~A., \& {Dello Russo}, N. 2003, \apj, 588, 535

\bibitem[{{Burke} \& {Hollenbach}(1983)}]{bur83}
{Burke}, J.~R. \& {Hollenbach}, D.~J. 1983, \apj, 265, 223

\bibitem[Carr \& Najita(2008)]{car08} Carr, J.~S., \& Najita, J.~R.\
2008, Science, 319, 1504

\bibitem[{{Carr} {et~al.}(2004){Carr}, {Tokunaga}, \& {Najita}}]{car04}
{Carr}, J.~S., {Tokunaga}, A.~T., \& {Najita}, J. 2004, \apj, 603, 213

\bibitem[Casassus et 
al.(2005)]{cas05} Casassus, S., Stahl, O., \& Wilson, T.~L.\ 2005, \aap, 441, 181 

\bibitem[{{Cazaux} \& {Tielens}(2002{\natexlab{a}})}]{caz02b}
{Cazaux}, S. \& {Tielens}, A.~G.~G.~M. 2002{\natexlab{a}}, \apjl, 577, L127

\bibitem[{{Cazaux} \& {Tielens}(2002{\natexlab{b}})}]{caz02a}
---. 2002{\natexlab{b}}, \apjl, 575, L29

\bibitem[{{Cazaux} \& {Tielens}(2004)}]{caz04}
---. 2004, \apj, 604, 222

\bibitem[Charnley et al.(2004)]{cha04} Charnley, S.~B., 
Ehrenfreund, P., Millar, T.~J., Boogert, A.~C.~A., Markwick, A.~J., Butner, 
H.~M., Ruiterkamp, R., \& Rodgers, S.~D.\ 2004, \mnras, 347, 157 

\bibitem[{{Chen} \& {Marcus}(2005)}]{che05}
{Chen}, W.-C. \& {Marcus}, R.~A. 2005, \jcp, 123

\bibitem[{{Clavel} {et~al.}(1978){Clavel}, {Viala}, \& {Bel}}]{cla78}
{Clavel}, J., {Viala}, Y.~P., \& {Bel}, N. 1978, \aap, 65, 435

\bibitem[{{Clayton} \& {Nittler}(2004)}]{cla04}
{Clayton}, D.~D. \& {Nittler}, L.~R. 2004, \araa, 42, 39

\bibitem[{{D'Alessio} {et~al.}(2001){D'Alessio}, {Calvet}, \&
  {Hartmann}}]{dal01}
{D'Alessio}, P., {Calvet}, N., \& {Hartmann}, L. 2001, \apj, 553, 321

\bibitem[{{D'Alessio} {et~al.}(2006){D'Alessio}, {Calvet}, {Hartmann},
  {Franco-Hern{\'a}ndez}, \& {Serv{\'{\i}}n}}]{dal06}
{D'Alessio}, P., {Calvet}, N., {Hartmann}, L., {Franco-Hern{\'a}ndez}, R., \&
  {Serv{\'{\i}}n}, H. 2006, \apj, 638, 314

\bibitem[{{D'Alessio} {et~al.}(1999){D'Alessio}, {Calvet}, {Hartmann},
  {Lizano}, \& {Cant{\'o}}}]{dal99}
{D'Alessio}, P., {Calvet}, N., {Hartmann}, L., {Lizano}, S., \& {Cant{\'o}}, J.
  1999, \apj, 527, 893

\bibitem[{{D'Alessio} {et~al.}(1998){D'Alessio}, {Canto}, {Calvet}, \&
  {Lizano}}]{dal98}
{D'Alessio}, P., {Canto}, J., {Calvet}, N., \& {Lizano}, S. 1998, \apj, 500,
  411

\bibitem[{{Dalgarno} \& {Black}(1976)}]{dal76}
{Dalgarno}, A. \& {Black}, J.~H. 1976, Reports of Progress in Physics, 39, 573

\bibitem[{{Dartois} {et~al.}(2003){Dartois}, {Dutrey}, \& {Guilloteau}}]{dar03}
{Dartois}, E., {Dutrey}, A., \& {Guilloteau}, S. 2003, \aap, 399, 773

\bibitem[{{de Jong} {et~al.}(1980){de Jong}, {Boland}, \& {Dalgarno}}]{dej80}
{de Jong}, T., {Boland}, W., \& {Dalgarno}, A. 1980, \aap, 91, 68

\bibitem[{{Desch} \& {Connolly}(2002)}]{des02}
{Desch}, S.~J. \& {Connolly}, Jr., H.~C. 2002, Meteoritics and Planetary
  Science, 37, 183

\bibitem[{{Draine} \& {Lee}(1984)}]{dra84}
{Draine}, B.~T. \& {Lee}, H.~M. 1984, \apj, 285, 89

\bibitem[{{Draine} {et~al.}(1983){Draine}, {Roberge}, \& {Dalgarno}}]{dra83}
{Draine}, B.~T., {Roberge}, W.~G., \& {Dalgarno}, A. 1983, \apj, 264, 485

\bibitem[{{Dutrey} {et~al.}(1997){Dutrey}, {Guilloteau}, \& {Guelin}}]{dut97}
{Dutrey}, A., {Guilloteau}, S., \& {Guelin}, M. 1997, \aap, 317, L55

\bibitem[Ehrenfreund \& Schutte(2000)]{ehr00}
  Ehrenfreund, P., \& Schutte, W.~A.\ 2000, From Molecular Clouds to
  Planetary, 197, 135

\bibitem[{{Eisenhauer} {et~al.}(2003){Eisenhauer}, {Sch{\"o}del}, {Genzel},
  {Ott}, {Tecza}, {Abuter}, {Eckart}, \& {Alexander}}]{eis03}
{Eisenhauer}, F., {Sch{\"o}del}, R., {Genzel}, R., {Ott}, T., {Tecza}, M.,
  {Abuter}, R., {Eckart}, A., \& {Alexander}, T. 2003, \apjl, 597, L121


\bibitem[{{Feigelson} {et~al.}(2002){Feigelson}, {Garmire}, \&
  {Pravdo}}]{fei02}
{Feigelson}, E.~D., {Garmire}, G.~P., \& {Pravdo}, S.~H. 2002, \apj, 572, 335

\bibitem[{{Feigelson} \& {Montmerle}(1999)}]{fei99}
{Feigelson}, E.~D. \& {Montmerle}, T. 1999, \araa, 37, 363

\bibitem[{{Fraser} {et~al.}(2001){Fraser}, {Collings}, {McCoustra}, \&
  {Williams}}]{fra01}
{Fraser}, H.~J., {Collings}, M.~P., {McCoustra}, M.~R.~S., \& {Williams}, D.~A.
  2001, \mnras, 327, 1165

\bibitem[{{Geers} {et~al.}(2006){Geers}, {Augereau}, {Pontoppidan},
  {Dullemond}, {Visser}, {Kessler-Silacci}, {Evans}, {van Dishoeck}, {Blake},
  {Boogert}, {Brown}, {Lahuis}, \& {Mer{\'{\i}}n}}]{gee06}
{Geers}, V.~C., {Augereau}, J.-C., {Pontoppidan}, K.~M., {Dullemond}, C.~P.,
  {Visser}, R., {Kessler-Silacci}, J.~E., {Evans}, II, N.~J., {van Dishoeck},
  E.~F., {Blake}, G.~A., {Boogert}, A.~C.~A., {Brown}, J.~M., {Lahuis}, F., \&
  {Mer{\'{\i}}n}, B. 2006, \aap, 459, 545

\bibitem[{{Gibb} {et~al.}(2004){Gibb}, {Rettig}, {Brittain}, {Haywood},
  {Simon}, \& {Kulesa}}]{gib04}
{Gibb}, E.~L., {Rettig}, T., {Brittain}, S., {Haywood}, R., {Simon}, T., \&
  {Kulesa}, C. 2004, \apjl, 610, L113

\bibitem[{{Gibb} {et~al.}(2007){Gibb}, {Van Brunt}, {Brittain}, \&
  {Rettig}}]{gib07}
{Gibb}, E.~L., {Van Brunt}, K.~A., {Brittain}, S.~D., \& {Rettig}, T.~W. 2007,
  \apj, 660, 1572

\bibitem[{{Gilman}(1972)}]{gil72}
{Gilman}, R.~C. 1972, \apj, 178, 423

\bibitem[{{Glassgold} \& {Langer}(1973)}]{gla73}
{Glassgold}, A.~E. \& {Langer}, W.~D. 1973, \apj, 186, 859

\bibitem[{{Glassgold} {et~al.}(2004){Glassgold}, {Najita}, \& {Igea}}]{gla04}
{Glassgold}, A.~E., {Najita}, J., \& {Igea}, J. 2004, \apj, 615, 972

\bibitem[{{Gorti} \& {Hollenbach}(2004)}]{gor04}
{Gorti}, U. \& {Hollenbach}, D. 2004, \apj, 613, 424

\bibitem[{{Graedel} {et~al.}(1982){Graedel}, {Langer}, \& {Frerking}}]{gra82}
{Graedel}, T.~E., {Langer}, W.~D., \& {Frerking}, M.~A. 1982, \apjs, 48, 321

\bibitem[Guilloteau \& Dutrey(1998)]{gui98} Guilloteau, S., \& Dutrey,
A.\ 1998, \aap, 339, 467

\bibitem[{{Gullbring} {et~al.}(1998){Gullbring}, {Hartmann}, {Briceno}, \&
  {Calvet}}]{gul98}
{Gullbring}, E., {Hartmann}, L., {Briceno}, C., \& {Calvet}, N. 1998, \apj,
  492, 323

\bibitem[{{Habing}(1968)}]{hab68}
{Habing}, H.~J. 1968, \bain, 19, 421

\bibitem[{{Hasegawa} \& {Herbst}(1993)}]{has93}
{Hasegawa}, T.~I. \& {Herbst}, E. 1993, \mnras, 261, 83

\bibitem[{{Hasegawa} {et~al.}(1992){Hasegawa}, {Herbst}, \& {Leung}}]{has92}
{Hasegawa}, T.~I., {Herbst}, E., \& {Leung}, C.~M. 1992, \apjs, 82, 167

\bibitem[{{Hayashi}(1981)}]{hay81}
{Hayashi}, C. 1981, Progress of Theoretical Physics Supplement, 70, 35

\bibitem[{{Hester} {et~al.}(2004){Hester}, {Desch}, {Healy}, \&
  {Leshin}}]{hes04}
{Hester}, J.~J., {Desch}, S.~J., {Healy}, K.~R., \& {Leshin}, L.~A. 2004,
  Science, 304, 1116

\bibitem[{{Hogerheijde} {et~al.}(1998){Hogerheijde}, {van Dishoeck}, {Blake},
  \& {van Langevelde}}]{hog98}
{Hogerheijde}, M.~R., {van Dishoeck}, E.~F., {Blake}, G.~A., \& {van
  Langevelde}, H.~J. 1998, \apj, 502, 315

\bibitem[{{Hollenbach} \& {McKee}(1979)}]{hol79}
{Hollenbach}, D. \& {McKee}, C.~F. 1979, \apjs, 41, 555

\bibitem[{{Hollenbach} {et~al.}(1991){Hollenbach}, {Takahashi}, \&
  {Tielens}}]{hol91}
{Hollenbach}, D.~J., {Takahashi}, T., \& {Tielens}, A.~G.~G.~M. 1991, \apj,
  377, 192

\bibitem[{{Iben} \& {Renzini}(1983)}]{ibe83}
{Iben}, Jr., I. \& {Renzini}, A. 1983, \araa, 21, 271

\bibitem[{{Igea} \& {Glassgold}(1999)}]{ige99}
{Igea}, J. \& {Glassgold}, A.~E. 1999, \apj, 518, 848

\bibitem[{{Ilgner} {et~al.}(2004){Ilgner}, {Henning}, {Markwick}, \&
  {Millar}}]{ilg04}
{Ilgner}, M., {Henning}, T., {Markwick}, A.~J., \& {Millar}, T.~J. 2004, \aap,
  415, 643

\bibitem[{{Ilgner} \& {Nelson}(2006)}]{ilg06}
{Ilgner}, M. \& {Nelson}, R.~P. 2006, \aap, 445, 205

\bibitem[Jewitt et al.(1997)]{jew97} Jewitt, D., Matthews, H.~E.,
Owen, T., \& Meier, R.\ 1997, Science, 278, 90

\bibitem[{{Kamp} \& {Dullemond}(2004)}]{kam04}
{Kamp}, I. \& {Dullemond}, C.~P. 2004, \apj, 615, 991

\bibitem[{{Kamp} \& {van Zadelhoff}(2001)}]{kam01}
{Kamp}, I. \& {van Zadelhoff}, G.-J. 2001, \aap, 373, 641

\bibitem[{{Kastner} {et~al.}(1997){Kastner}, {Zuckerman}, {Weintraub}, \&
  {Forveille}}]{kas97}
{Kastner}, J.~H., {Zuckerman}, B., {Weintraub}, D.~A., \& {Forveille}, T. 1997,
  Science, 277, 67

\bibitem[{{Keene} {et~al.}(1998){Keene}, {Schilke}, {Kooi}, {Lis}, {Mehringer},
  \& {Phillips}}]{kee98}
{Keene}, J., {Schilke}, P., {Kooi}, J., {Lis}, D.~C., {Mehringer}, D.~M., \&
  {Phillips}, T.~G. 1998, \apjl, 494, L107+

\bibitem[{{Kress} {et~al.}(2002){Kress}, {Desch}, {Dateo}, \&
  {Benedix}}]{kre02}
{Kress}, M.~E., {Desch}, S.~J., {Dateo}, C.~E., \& {Benedix}, G. 2002, Advances
  in Space Research, 30, 1473

\bibitem[{{Lahuis} {et~al.}(2006){Lahuis}, {van Dishoeck}, {Boogert},
  {Pontoppidan}, {Blake}, {Dullemond}, {Evans}, {Hogerheijde}, {J{\o}rgensen},
  {Kessler-Silacci}, \& {Knez}}]{lah06}
{Lahuis}, F., {van Dishoeck}, E.~F., {Boogert}, A.~C.~A., {Pontoppidan}, K.~M.,
  {Blake}, G.~A., {Dullemond}, C.~P., {Evans}, II, N.~J., {Hogerheijde}, M.~R.,
  {J{\o}rgensen}, J.~K., {Kessler-Silacci}, J.~E., \& {Knez}, C. 2006, \apjl,
  636, L145

\bibitem[{{Langer}(1992)}]{lan92}
{Langer}, W.~D. 1992, in IAU Symposium, Vol. 150, Astrochemistry of Cosmic
  Phenomena, ed. P.~D. {Singh}, 193--+

\bibitem[Langer \& Penzias(1990)]{lan90} Langer, W.~D., \& Penzias,
A.~A.\ 1990, \apj, 357, 477

\bibitem[{{Langer} \& {Graedel}(1989)}]{lan89}
{Langer}, W.~D. \& {Graedel}, T.~E. 1989, \apjs, 69, 241

\bibitem[{{Langer} {et~al.}(1984){Langer}, {Graedel}, {Frerking}, \&
  {Armentrout}}]{lan84}
{Langer}, W.~D., {Graedel}, T.~E., {Frerking}, M.~A., \& {Armentrout}, P.~B.
  1984, \apj, 277, 581

\bibitem[{{Le Bourlot} {et~al.}(1999){Le Bourlot}, {Pineau des For{\^e}ts}, \&
  {Flower}}]{leb99}
{Le Bourlot}, J., {Pineau des For{\^e}ts}, G., \& {Flower}, D.~R. 1999, \mnras,
  305, 802

\bibitem[{{Le Teuff} {et~al.}(2000){Le Teuff}, {Millar}, \& {Markwick}}]{let00}
{Le Teuff}, Y.~H., {Millar}, T.~J., \& {Markwick}, A.~J. 2000, \aaps, 146, 157

\bibitem[{{Lecluse} {et~al.}(1998){Lecluse}, {Robert}, {Kaiser}, {Roessler},
  {Pillinger}, \& {Javoy}}]{lec98}
{Lecluse}, C., {Robert}, F., {Kaiser}, R.-I., {Roessler}, K., {Pillinger},
  C.~T., \& {Javoy}, M. 1998, \aap, 330, 1175

\bibitem[{{Lee} {et~al.}(1996){Lee}, {Herbst}, {Pineau des Forets}, {Roueff},
  \& {Le Bourlot}}]{lee96}
{Lee}, H.-H., {Herbst}, E., {Pineau des Forets}, G., {Roueff}, E., \& {Le
  Bourlot}, J. 1996, \aap, 311, 690

\bibitem[Lis et al.(1997)]{lis97} Lis, D.~C., et al.\ 1997, Icarus,
130, 355

\bibitem[Liszt(2007)]{lis07} Liszt, H.~S.\ 2007, \aap, 476, 291 

\bibitem[{{Lohr}(1998)}]{loh98}
{Lohr}, L.~L. 1998, \jcp, 108, 8012

\bibitem[{{London}(1978)}]{lon78}
{London}, R. 1978, \apj, 225, 405

\bibitem[{{Lyons} {et~al.}(2007){Lyons}, {Boney}, \& {Marcus}}]{lyo07}
{Lyons}, J.~R., {Boney}, E., \& {Marcus}, R.~A. 2007, in Lunar and Planetary
  Institute Conference Abstracts, Vol.~38, Lunar and Planetary Institute
  Conference Abstracts, 2382--+

\bibitem[{{Maloney} {et~al.}(1996){Maloney}, {Hollenbach}, \&
  {Tielens}}]{mal96}
{Maloney}, P.~R., {Hollenbach}, D.~J., \& {Tielens}, A.~G.~G.~M. 1996, \apj,
  466, 561

\bibitem[{{Markwick} {et~al.}(2002){Markwick}, {Ilgner}, {Millar}, \&
  {Henning}}]{mar02}
{Markwick}, A.~J., {Ilgner}, M., {Millar}, T.~J., \& {Henning}, T. 2002, \aap,
  385, 632


\bibitem[{{Meibom} {et~al.}(2007){Meibom}, {Krot}, {Robert}, {Mostefaoui},
  {Russell}, {Petaev}, \& {Gounelle}}]{mei07}
{Meibom}, A., {Krot}, A.~N., {Robert}, F., {Mostefaoui}, S., {Russell}, S.~S.,
  {Petaev}, M.~I., \& {Gounelle}, M. 2007, \apjl, 656, L33

\bibitem[{{Messenger}(2000)}]{mes00}
{Messenger}, S. 2000, \nat, 404, 968

\bibitem[{{Milam} {et~al.}(2005){Milam}, {Savage}, {Brewster}, {Ziurys}, \&
  {Wyckoff}}]{mil05}
{Milam}, S.~N., {Savage}, C., {Brewster}, M.~A., {Ziurys}, L.~M., \& {Wyckoff},
  S. 2005, \apj, 634, 1126

\bibitem[{{Millar} {et~al.}(2003){Millar}, {Nomura}, \& {Markwick}}]{mil03}
{Millar}, T.~J., {Nomura}, H., \& {Markwick}, A.~J. 2003, \apss, 285, 761

\bibitem[{{Najita} {et~al.}(2003){Najita}, {Carr}, \& {Mathieu}}]{naj03}
{Najita}, J., {Carr}, J.~S., \& {Mathieu}, R.~D. 2003, \apj, 589, 931

\bibitem[{{Nomura}(2002)}]{nom02}
{Nomura}, H. 2002, \apj, 567, 587

\bibitem[{{{\"O}berg} {et~al.}(2005){{\"O}berg}, {van Broekhuizen}, {Fraser},
  {Bisschop}, {van Dishoeck}, \& {Schlemmer}}]{obe05}
{{\"O}berg}, K.~I., {van Broekhuizen}, F., {Fraser}, H.~J., {Bisschop}, S.~E.,
  {van Dishoeck}, E.~F., \& {Schlemmer}, S. 2005, \apjl, 621, L33

\bibitem[{{Pi{\'e}tu} {et~al.}(2007){Pi{\'e}tu}, {Dutrey}, \&
  {Guilloteau}}]{pie07}
{Pi{\'e}tu}, V., {Dutrey}, A., \& {Guilloteau}, S. 2007, \aap, 467, 163

\bibitem[Preibisch et al.(1993)]{pre93} Preibisch, T., Ossenkopf, V.,
Yorke, H.~W., \& Henning, T.\ 1993, \aap, 279, 577

\bibitem[{{Press} {et~al.}(1992){Press}, {Teukolsky}, {Vetterling}, \&
  {Flannery}}]{pre92}
{Press}, W.~H., {Teukolsky}, S.~A., {Vetterling}, W.~T., \& {Flannery}, B.~P.
  1992, {Numerical recipes in FORTRAN. The art of scientific computing}
  (Cambridge: University Press, |c1992, 2nd ed.)

\bibitem[{{Qi} {et~al.}(2003){Qi}, {Kessler}, {Koerner}, {Sargent}, \&
  {Blake}}]{qi03}
{Qi}, C., {Kessler}, J.~E., {Koerner}, D.~W., {Sargent}, A.~I., \& {Blake},
  G.~A. 2003, \apj, 597, 986

\bibitem[Richling \& Yorke(2000)]{ric00} Richling, S., \& Yorke,
   H.~W.\ 2000, \apj, 539, 258

\bibitem[{{Roeckmann} {et~al.}(1998){Roeckmann}, {Brenninkmeijer},
  {Saueressig}, {Bergamaschi}, {Crowley}, {Fischer}, \& {Crutzen}}]{roc98}
{Roeckmann}, T., {Brenninkmeijer}, C.~A.~M., {Saueressig}, G., {Bergamaschi},
  P., {Crowley}, J.~N., {Fischer}, H., \& {Crutzen}, P.~J. 1998, Science, 281,
  544

\bibitem[Sakai et al.(2007)]{sak07} Sakai, N., Ikeda, M., 
Morita, M., Sakai, T., Takano, S., Osamura, Y., 
\& Yamamoto, S.\ 2007, \apj, 663, 1174 

\bibitem[{{Schinke} {et~al.}(1985){Schinke}, {Engel}, {Buck}, {Meyer}, \&
  {Diercksen}}]{sch85}
{Schinke}, R., {Engel}, V., {Buck}, U., {Meyer}, H., \& {Diercksen}, G.~H.~F.
  1985, \apj, 299, 939

\bibitem[{{Semenov} {et~al.}(2005){Semenov}, {Pavlyuchenkov}, {Schreyer},
  {Henning}, {Dullemond}, \& {Bacmann}}]{sem05}
{Semenov}, D., {Pavlyuchenkov}, Y., {Schreyer}, K., {Henning}, T., {Dullemond},
  C., \& {Bacmann}, A. 2005, \apj, 621, 853

\bibitem[{{Semenov} {et~al.}(2006){Semenov}, {Wiebe}, \& {Henning}}]{sem06}
{Semenov}, D., {Wiebe}, D., \& {Henning}, T. 2006, \apjl, 647, L57

\bibitem[{{Shakura} \& {Syunyaev}(1973)}]{sha73}
{Shakura}, N.~I. \& {Syunyaev}, R.~A. 1973, \aap, 24, 337

\bibitem[{{Shang} {et~al.}(2002){Shang}, {Glassgold}, {Shu}, \&
  {Lizano}}]{sha02}
{Shang}, H., {Glassgold}, A.~E., {Shu}, F.~H., \& {Lizano}, S. 2002, \apj, 564,
  853

\bibitem[{{Sheffer} {et~al.}(1992){Sheffer}, {Federman}, {Lambert}, \&
  {Cardelli}}]{she92}
{Sheffer}, Y., {Federman}, S.~R., {Lambert}, D.~L., \& {Cardelli}, J.~A. 1992,
  \apj, 397, 482

\bibitem[{{Smit} {et~al.}(1982){Smit}, {Volz}, {Ehhalt}, \& {Knappe}}]{smi82}
{Smit}, H.~G.~J., {Volz}, A., {Ehhalt}, D.~H., \& {Knappe}, H. 1982, in Stable
  Isotopes, ed. H.-L. {Schmidt}, H.~{F\"orstel}, \& K.~{Heizinger} (Amsterdam:
  Elsevier, 1982), 147--152

\bibitem[{{Smith} \& {Adams}(1980)}]{smi80}
{Smith}, D. \& {Adams}, N.~G. 1980, \apj, 242, 424

\bibitem[{{Stephens} {et~al.}(1980){Stephens}, {Kaplan}, {Gorse}, {Durkee},
  {Compton}, {Cohen}, \& {Bielling}}]{ste80}
{Stephens}, C.~M., {Kaplan}, L., {Gorse}, R., {Durkee}, S., {Compton}, M.,
  {Cohen}, S., \& {Bielling}, K. 1980, ijck, 12, 935

\bibitem[{{Stephens} \& {Dalgarno}(1973)}]{ste73}
{Stephens}, T.~L. \& {Dalgarno}, A. 1973, \apj, 186, 165

\bibitem[{{Sternberg} \& {Dalgarno}(1989)}]{ste89}
{Sternberg}, A. \& {Dalgarno}, A. 1989, \apj, 338, 197

\bibitem[Takano et al.(1998)]{tak98} Takano, S., et al.\ 1998, \aap,
329, 1156

\bibitem[{{Thi} {et~al.}(2004){Thi}, {van Zadelhoff}, \& {van
  Dishoeck}}]{thi04}
{Thi}, W.-F., {van Zadelhoff}, G.-J., \& {van Dishoeck}, E.~F. 2004, \aap, 425,
  955

\bibitem[{{Tielens}(1983)}]{tie83}
{Tielens}, A.~G.~G.~M. 1983, \apj, 271, 702

\bibitem[{{Tielens} \& {Hollenbach}(1985)}]{tie85}
{Tielens}, A.~G.~G.~M. \& {Hollenbach}, D. 1985, \apj, 291, 722

\bibitem[{{Timmes} {et~al.}(1995){Timmes}, {Woosley}, \& {Weaver}}]{tim95}
{Timmes}, F.~X., {Woosley}, S.~E., \& {Weaver}, T.~A. 1995, \apjs, 98, 617

\bibitem[{{Tosi}(1982)}]{tos82}
{Tosi}, M. 1982, \apj, 254, 699

\bibitem[{{Turner} {et~al.}(2007){Turner}, {Sano}, \& {Dziourkevitch}}]{tur07}
{Turner}, N.~J., {Sano}, T., \& {Dziourkevitch}, N. 2007, \apj, 659, 729

\bibitem[{{Umebayashi} \& {Nakano}(1981)}]{ume81}
{Umebayashi}, T. \& {Nakano}, T. 1981, \pasj, 33, 617

\bibitem[{{van Dishoeck} \& {Black}(1988)}]{van88}
{van Dishoeck}, E.~F. \& {Black}, J.~H. 1988, \apj, 334, 771

\bibitem[{{van Zadelhoff} {et~al.}(2003){van Zadelhoff}, {Aikawa},
  {Hogerheijde}, \& {van Dishoeck}}]{van03}
{van Zadelhoff}, G.-J., {Aikawa}, Y., {Hogerheijde}, M.~R., \& {van Dishoeck},
  E.~F. 2003, \aap, 397, 789

\bibitem[Verner \& Yakovlev(1995)]{ver95} Verner, D.~A., \& Yakovlev,
D.~G.\ 1995, \aaps, 109, 125

\bibitem[{{Watson} {et~al.}(1976){Watson}, {Anicich}, \& {Huntress}}]{wat76}
{Watson}, W.~D., {Anicich}, V.~G., \& {Huntress}, Jr., W.~T. 1976, \apjl, 205,
  L165

\bibitem[{{Werner}(1970)}]{wer70}
{Werner}, M.~W. 1970, \aplett, 6, 81

\bibitem[{{White} \& {Hillenbrand}(2004)}]{whi04}
{White}, R.~J. \& {Hillenbrand}, L.~A. 2004, \apj, 616, 998

\bibitem[{{Wielen} {et~al.}(1996){Wielen}, {Fuchs}, \& {Dettbarn}}]{wie96}
{Wielen}, R., {Fuchs}, B., \& {Dettbarn}, C. 1996, \aap, 314, 438

\bibitem[{{Wielen} \& {Wilson}(1997)}]{wie97}
{Wielen}, R. \& {Wilson}, T.~L. 1997, \aap, 326, 139

\bibitem[{{Willacy} {et~al.}(2006){Willacy}, {Langer}, {Allen}, \&
  {Bryden}}]{wil06}
{Willacy}, K., {Langer}, W., {Allen}, M., \& {Bryden}, G. 2006, \apj, 644, 1202

\bibitem[{{Wilms} {et~al.}(2000){Wilms}, {Allen}, \& {McCray}}]{wilms00}
{Wilms}, J., {Allen}, A., \& {McCray}, R. 2000, \apj, 542, 914

\bibitem[{{Wilson} \& {Rood}(1994)}]{wil94}
{Wilson}, T.~L. \& {Rood}, R. 1994, \araa, 32, 191

\bibitem[{{Woodall} {et~al.}(2007){Woodall}, {Ag{\'u}ndez}, {Markwick-Kemper},
  \& {Millar}}]{umist06}
{Woodall}, J., {Ag{\'u}ndez}, M., {Markwick-Kemper}, A.~J., \& {Millar}, T.~J.
  2007, \aap, 466, 1197

\bibitem[{{Woods} \& {Willacy}(2007)}]{woo07}
{Woods}, P.~M. \& {Willacy}, K. 2007, \apjl, 655, L49


\bibitem[{{Young}(2006)}]{you06}
{Young}, E.~D. 2006, in Lunar and Planetary Institute Conference Abstracts,
  Vol.~37, 37th Annual Lunar and Planetary Science Conference, ed.
  S.~{Mackwell} \& E.~{Stansbery}, 1790--+

\bibitem[Ziurys et al.(1999)]{ziu99} Ziurys, L.~M., Savage, C.,
Brewster, M.~A., Apponi, A.~J., Pesch, T.~C., \& Wyckoff, S.\ 1999,
\apjl, 527, L67

\end{thebibliography}
\end{document}